\documentstyle{mn}

\newif\ifAMStwofonts

\ifoldfss
  \ifCUPmtlplainloaded \else
    \NewTextAlphabet{textbfit} {cmbxti10} {}
    \NewTextAlphabet{textbfss} {cmssbx10} {}
    \NewMathAlphabet{mathbfit} {cmbxti10} {} 
    \NewMathAlphabet{mathbfss} {cmssbx10} {} 
  \fi
  \ifAMStwofonts
    \ifCUPmtlplainloaded \else
      \NewSymbolFont{upmath} {eurm10}
      \NewSymbolFont{AMSa} {msam10}
      \NewMathSymbol{\upi}     {0}{upmath}{19}
      \NewMathSymbol{\umu}     {0}{upmath}{16}
      \NewMathSymbol{\upartial}{0}{upmath}{40}
      \NewMathSymbol{\leqslant}{3}{AMSa}{36}
      \NewMathSymbol{\geqslant}{3}{AMSa}{3E}

      \let\leq=\leqslant \let\le=\leqslant
       \let\ge=\geqslant
    \fi
  \fi
\fi 

\ifnfssone
  \newmathalphabet{\mathit}
  \addtoversion{normal}{\mathit}{cmr}{m}{it}
  \addtoversion{bold}{\mathit}{cmr}{bx}{it}
  \newmathalphabet{\mathbfit} 
  \addtoversion{normal}{\mathbfit}{cmr}{bx}{it}
  \addtoversion{bold}{\mathbfit}{cmr}{bx}{it}
  \newmathalphabet{\mathbfss} 
  \addtoversion{normal}{\mathbfss}{cmss}{bx}{n}
  \addtoversion{bold}{\mathbfss}{cmss}{bx}{n}
  \ifAMStwofonts
    \ifCUPmtlplainloaded \else
      \UseAMStwoboldmath
      \makeatletter
      \new@mathgroup\upmath@group
      \define@mathgroup\mv@normal\upmath@group{eur}{m}{n}
      \define@mathgroup\mv@bold\upmath@group{eur}{b}{n}
      \edef\UPM{\hexnumber\upmath@group}
      \new@mathgroup\amsa@group
      \define@mathgroup\mv@normal\amsa@group{msa}{m}{n}
      \define@mathgroup\mv@bold\amsa@group{msa}{m}{n}
      \edef\AMSa{\hexnumber\amsa@group}
      \makeatother
      \mathchardef\upi="0\UPM19
      \mathchardef\umu="0\UPM16
      \mathchardef\upartial="0\UPM40
      \mathchardef\leqslant="3\AMSa36
      \mathchardef\geqslant="3\AMSa3E

      \let\leq=\leqslant \let\le=\leqslant
       \let\ge=\geqslant
    \fi
  \fi
\fi 

\ifnfsstwo
  \DeclareMathAlphabet{\mathbfit}{OT1}{cmr}{bx}{it}
  \SetMathAlphabet\mathbfit{bold}{OT1}{cmr}{bx}{it}
  \DeclareMathAlphabet{\mathbfss}{OT1}{cmss}{bx}{n}
  \SetMathAlphabet\mathbfss{bold}{OT1}{cmss}{bx}{n}
  \ifAMStwofonts
    \ifCUPmtlplainloaded \else
      \DeclareSymbolFont{UPM}{U}{eur}{m}{n}
      \SetSymbolFont{UPM}{bold}{U}{eur}{b}{n}
      \DeclareSymbolFont{AMSa}{U}{msa}{m}{n}
      \DeclareMathSymbol{\upi}{0}{UPM}{"19}
      \DeclareMathSymbol{\umu}{0}{UPM}{"16}
      \DeclareMathSymbol{\upartial}{0}{UPM}{"40}
      \DeclareMathSymbol{\leqslant}{3}{AMSa}{"36}
      \DeclareMathSymbol{\geqslant}{3}{AMSa}{"3E}

      \let\leq=\leqslant \let\le=\leqslant
       \let\ge=\geqslant
    \fi
  \fi
\fi 

\ifCUPmtlplainloaded \else
  \ifAMStwofonts \else 
    \def\upi{\pi}
    \def\umu{\mu}
    \def\upartial{\partial}
  \fi
\fi

\title{Spatially extended absorption around the z$=$2.63 radio galaxy MRC 2025-218: outflow or infall?}
\author[Humphrey et al.]
       {A. Humphrey$^{1,2}$, M. Villar-Mart\'\i n$^{3}$, S. F. S\'anchez$^{4}$, S. di Serego Alighieri$^{5}$, \newauthor C. De Breuck$^{6}$, L. Binette$^{1,7}$, C. Tadhunter$^{8}$, J. Vernet$^{6}$, R. Fosbury$^{9}$, J. Stasielak$^{2}$\\
$^{1}$Instituto de Astronom\'\i a, Universidad Nacional Aut\'onomo de M\'exico, Ap. 70-264, 04510 M\'exico, DF, M\'exico\\
$^{2}$Korea Astronomy and Space Science Institute, 61-1 Hwaam-dong, Yuseong-gu, Daejeon, 305-348, Republic of Korea (ajh@kasi.re.kr)\\
$^{3}$Instituto de Astrof\'\i sica de Andaluc\'\i a (CSIC), Aptdo. 3004, 18080 Granada, Spain\\
$^{4}$Centro Astron\'omico Hispano Alem\'an, Calar Alto, CSIC-MPG, C/Jes\'us Durb\'an Rem\'on 2-2, 04004 Almeria, Spain\\
$^{5}$INAF-Osservatorio Astrofisico di Arcetri, Largo Enrico Fermi 5, I-50125 Firenze, Italy\\
$^{6}$European Southern Observatory, Karl-Schwarschild Strasse, 85748 Garching bei M\"unchen, Germany\\
$^{7}$D\'{e}partement de Physique, de G\'{e}nie Physique et d'Optique, Universit\'{e} Laval, Qu\'{e}bec, QC, G1K\,7P4, Canada\\
$^{8}$Department of Physics \& Astronomy, University of Sheffield, Sheffield S3 7RH, United Kingdom\\
$^{9}$Space Telescope - European Coordination Facility, Karl-Schwarzschild Strasse 2, 85748 Garching-bei-M\"unchen, Germany}

\date{Accepted 12 Aug 2008.
      Received 7 Aug 2008;
      in original form 13 Jun 2008}

\pagerange{\pageref{firstpage}--\pageref{lastpage}}
\pubyear{2008}

\begin{document}

\maketitle

\label{firstpage}

\begin{abstract}
We present an investigation into the absorber in front of the z$=$2.63 radio galaxy MRC 2025-218, using integral field spectroscopy obtained at the Very Large Telescope, and long slit spectroscopy obtained at the Keck II telescope.  The properties of MRC 2025-218 are particularly conducive to study the nature of the absorbing gas, i.e., this galaxy shows bright and spatially extended Ly$\alpha$ emission, along with bright continuum emission from the active nucleus.  

Ly$\alpha$ absorption is detected across $\sim 40\times30$ kpc$^{2}$, has a covering factor of $\sim$1, and shows remarkably little variation in its properties across its entire spatial extent.  This absorber is kinematically detached from the extended emission line region (EELR).  Its properties suggest that the absorber is outside of the EELR.  We derive lower limits to the HI, HII and H column densities for this absorber of $3 \times 10^{16}$, $7 \times 10^{17}$ and $2 \times 10^{18}$ cm$^{-2}$, respectively.  Moreover, the relatively bright emission from the active nucleus has allowed us to measure a number of metal absorption lines: CI, CII, CIV, NV, OI, SiII, SiIV, AlII and AlIII.  The column density ratios are most naturally explained using photoionization by a hard continuum, with an ionization parameter U$\sim$0.0005-0.005.  Shocks or photoionization by young stars cannot reproduce satisfactorily the measured column ratios.  Using the ratio between the SiII* and SiII column densities, we derive a lower limit of $\ge$10 cm$^{-3}$ for the electron density of the absorber.  The data do not allow useful constraints to be placed on the metallicity of the absorber.  

We consider two possibilities for the nature of this absorber: the cosmological infall of gas, and an outflow driven by supernovae or the radio-jets.  We find it plausible that the absorber around 2025-218 is in outflow.  We also find good agreement between the observed properties of the HI absorber and the properties of the HI absorption expected from the cosmological infall model of Barkana \& Loeb.  

\end{abstract}


\section{Introduction}
Powerful active galaxies are important probes for understanding the formation and evolution of massive ellipticals and clusters of galaxies (e.g. Rocca-Volmerange et al. 2004; Kurk et al. 2000).  Their prodigious luminosities allow the identification of large samples spanning a vast range in z (e.g. R\"ottgering et al. 1997).  

Many powerful radio galaxies at z$>$2 (HzRG) are embedded within giant nebulae of ionized gas which strongly emit both metal lines and the recombination lines of H and He (e.g. McCarthy et al. 1987; Villar-Mart\'\i n et al. 2003).  In addition, they often show strong absorption features in Ly$\alpha$ (e.g. van Ojik et al. 1997; Wilman et al. 2004), and sometimes in metal lines such as CIV (e.g., Binette et al. 2000; Jarvis et al. 2003).  These absorption lines are thought to be formed in cold/warm gas inbetween the emission source and the observer, and as such they can provide a wealth of information about the active nucleus, the host galaxy and its environment (e.g. Hamann \& Ferland 1999).  They are often spatially very extended (e.g. R\"ottgering et al. 1995).  In their high spectral resolution study of the Ly$\alpha$ profile of 15 HzRG, van Ojik et al. (1997) noted: (i) an anticorrelation between the HI column of the absorber and the size of the radio source; (ii) the HI absorption features tend to be blueshifted relative to the line emission; and (iii) HI absorption is more frequently detected in smaller radio sources.  This has led various authors to suggest that the absorbers are part of an outflow, powered by either the radio source (Krause 2002; Wilman et al. 2004) or a massive starburst (Binette et al. 2000).  In at least some HzRG, the absorbers are enriched in metals, although high metallicities are not necessarily required (Binette et al. 2006).  Studying these absorbers is clearly important for understanding the formation and evolution of the host galaxy, and the possible symbiosis between this process and the nuclear activity.  

The high-z radio galaxy MRC 2025-218 (z=2.63: McCarthy et al. 1996) has been extensively studied in a variety of wavebands.  Radio maps observed at 5 and 8 GHz (17 and 30 GHz in the rest-frame; Carilli et al. 1997) show that MRC 2025-218 has a relatively small ($\sim$40 kpc), triple radio source, with a jet being clearly detected on one side of the nucleus.  Rest-frame UV and optical {\it Hubble Space Telescope (HST hereinafter)} images obtained by Pentericci et al. (1999; 2001) show a bright point source -- the active nucleus -- surrounded by relatively fainter, spatially extended emission from the host galaxy.  The UV continuum has a high degree of linear polarization (8.3$\pm$2.3 per cent), suggesting that a substantial fraction of this emission is scattered nuclear light (Cimatti et al. 1994).  In addition, UV-optical spectroscopy has revealed the existence of a giant nebula of warm ionized gas, with a spatial extent exceeding that of the radio source (Villar-Mart\'\i n et al. 1999, 2007; Pentericci et al. 2001; Humphrey et al. 2008).  This nebula emits an emission line spectrum characteristic of the low-density, extended emission line regions (EELR hereinafter) of warm ionized gas that are commonly associated with powerful active galaxies.  Very broad CIV and H$\alpha$ emission from the broad line region (BLR hereinafter) is also detected (Villar-Mart\'\i n et al. 1999; Larkin et al. 2000; Humphrey et al. 2008).  Based on {\it Spitzer} photometry of MRC 2025-218, Seymour et al. (2007) have derived a upper limit to the stellar mass of $\la$10$^{11.6} M_{odot}$.

Several properties of MRC 2025-218 are particularly conducive to the detection and study of intervening absorbers.  Firstly, the Ly$\alpha$ emission from the EELR is bright and has a large spatial extent, allowing HI absorption to be traced across large spatial scales.  Secondly, the brightness of the UV continuum emission, due to a strong contribution from the active nucleus itself, increases the probability of detecting lines in absorption that are usually weak or absent in emission in radio galaxies.  In a recent paper (Villar-Mart\'\i n et al. 2007: Paper I) we reported the detection of a strong, spatially extended Ly$\alpha$ absorption feature in the spectrum of MRC 2025-218.  Absorption from SiII, SiIV, OI, CII and CIV have also been detected in the spectrum of this source (Villar-Mart\'\i n et al. 1999).  In this paper we present a more detailed analysis of the absorbing gas in front of MRC 2025-218, making use of integral field spectroscopy from the Very Large Telecope (Paper I) and a deep, long slit spectrum from the Keck II telescope (Humphrey et al. 2008).  We also examine whether the absorbing gas might be related to an outflow of gas from the host galaxy, or the cosmological infall of gas onto the host galaxy.  Throughout this paper, we assume a flat universe with $H_{0}$=71 km $s^{-1}$ Mpc$^{-1}$, $\Omega_{\Lambda}$=0.73 and $\Omega_{m}$=0.27.  At the redshift of this source, 1\arcsec corresponds to 8.1 kpc.  

\begin{figure*}
\includegraphics{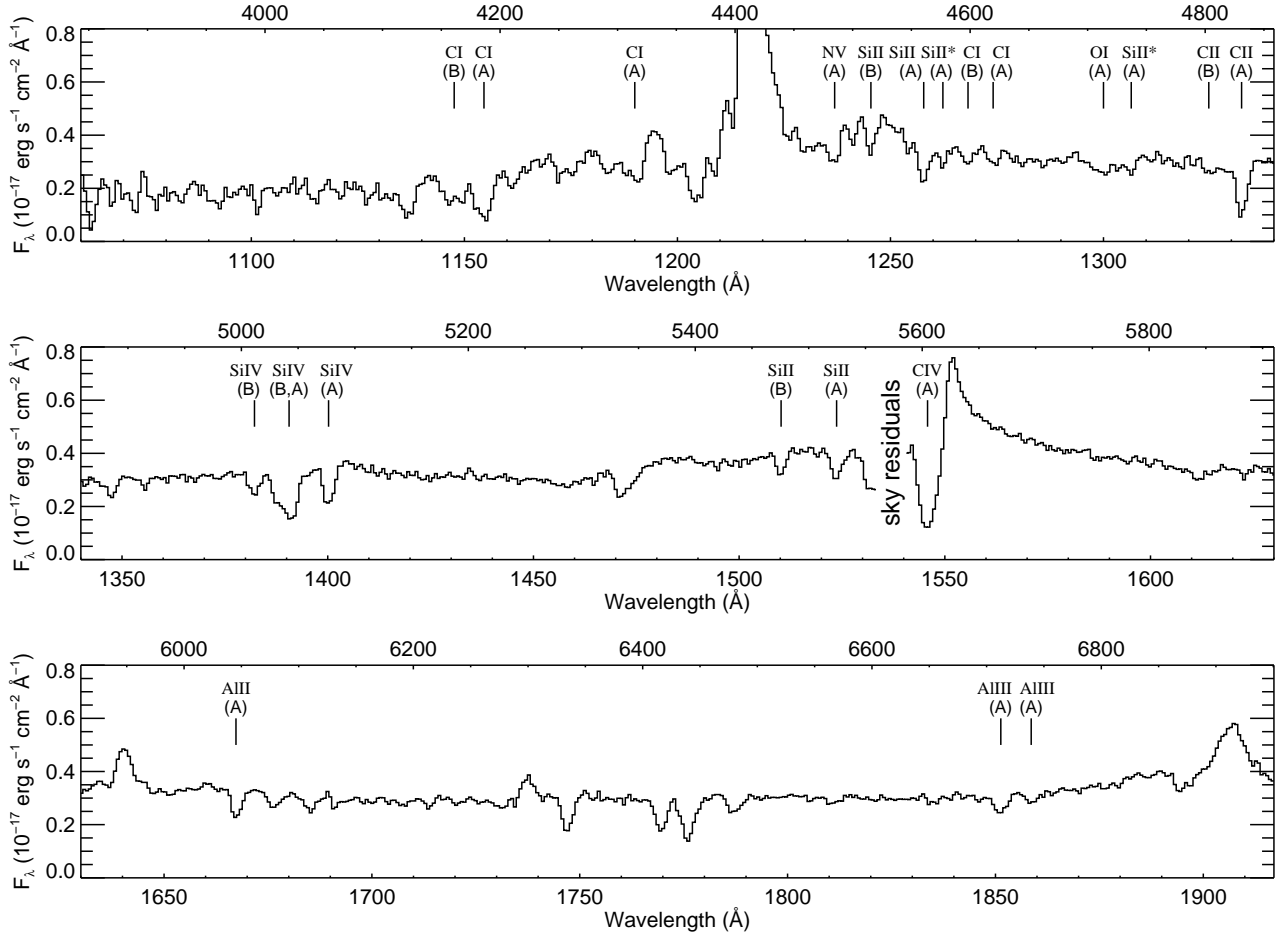}\vspace{5.1in}
\caption{The Keck II LRIS spectrum of MRC 2025-218, extracted from a 1.9\arcsec$\times$1.0\arcsec aperture.  The x-axes show both the observed and the rest-frame wavelength scales.  Identifications of the absorption lines associated with MRC 2025-218 are given.  We detect two absorption line systems: Absorber A, with a blueshift of several hundred km s$^{-1}$ relative to the narrow HeII $\lambda$1640 emission line; and absorber B, with a blueshift of a few thousand km s$^{-1}$ relative to the narrow HeII emission (see text).  Unmarked absorption features have ambiguous identifications or are not associated with the radio galaxy.}
\end{figure*}

\section{Data}

\subsection{Keck II LRIS observations}

Observations of MRC 2025-218 were made on 13 and 14 July 1999 at the Keck II telescope using the Low Resolution Imaging Spectrometer (LRIS: Oke et al. 1995) in polarimetry mode (Goodrich et al. 1995).  The LRIS detector has a spatial scale of 0.214\arcsec per pixel.  The slit was 1\arcsec wide and oriented at a position angle of 175$^{\circ}$, i.e., along the main axis of the radio emission (Carilli et al. 1997).  The full width at half of maximum (FWHM hereinafter) of the seeing disc was $\sim$1\arcsec, and thus the source filled the slit.  A 300 line mm$^{-1}$ grating was used, providing a dispersion of 2.4 \AA~per pixel, an effective spectral resolution (FWHM) of $\sim$10 \AA, and a spectral range of $\sim$3800-9000 \AA, or $\sim$1060-2430 \AA~in the rest-frame.  The source was observed for a total of 5.1 hr.  The observations were reduced using standard routines (see Cimatti et al. 1998; Vernet et al. 2001 for further details).  

\subsection{VLT VIMOS observations}

In addition, MRC 2025-218 was observed on 2005 July 28, 29 and 30 using the VIsible MultiObject Spectrograph (VIMOS: Le F\'evre et al. 2003), at the UT3 unit of the Very Large Telescope (VLT).  The integral field unit (IFU) of the VIMOS instrument comprises 6400 microlenses coupled to fibres, 1600 of which are used when in high resolution mode.  In this mode, the IFU covers 27\arcsec x 27\arcsec on the sky and each fibre has a diameter of 0.67\arcsec.  The HR$_{\rm blue}$ grating was used, resulting in an effective wavelength range of $\sim$4150-6200 \AA~ and an instrumental profile with a full width at half maximum (FWHM) of 1.7$\pm$0.2 \AA.  The total on-object exposure time was 10 hr.  During the observations, the FWHM of the seeing disc ranged from $\sim$0.4\arcsec to 3.0\arcsec.  The FWHM of the final co-added data cube, measured from a bright field star, is $\sim$1\arcsec.  

The data were reduced using R3D (S\'anchez 2006) and IRAF routines.  The original wavelength calibration was unreliable, and so the wavelength calibration was refined using the Keck II spectrum.  In the 4650-5250 \AA~ spectral region we were unable to achieve an accurate wavelength calibration, due to a lack of arc emission lines, sky lines or strong lines from the radio galaxy itself.  See Paper I for a more detailed description of the observations and a discussion of the emission line nebula associated with MRC 2025-218.

\section{Results}
\subsection{Keck II LRIS spectrum}
In the LRIS spectrum, we identify two separate absorbers associated with MRC 2025-218.  One is blueshifted by $\sim$600 km s$^{-1}$ relative to the velocity centroid of HeII $\lambda$1640, and is clearly detected in CI $\lambda$1157, CI $\lambda$1194, CI $\lambda$1278, CII $\lambda$1335, CIV $\lambda\lambda$1548,1551, NV $\lambda$1239, OI $\lambda$1302, SiII $\lambda$1260, SiII* $\lambda$1265, SiII* $\lambda$1309, SiII $\lambda$1527, SiIV $\lambda$1403, AlII $\lambda$1671 and AlIII $\lambda\lambda$1855,1863 (absorber A hereinafter).  Since these absorption lines all have a similar line of sight velocity, within the 1$\sigma$ uncertainties, we assume that they originate from the same absorber.  

The other absorber (absorber B) is blueshifted by $\sim$2000 km s$^{-1}$ relative to HeII, is detected in CI $\lambda$1157, CI $\lambda$1278, CII $\lambda$1335, and SiIV $\lambda$1394, and has lower equivalent width (W$_{\lambda}$) than does absorber A.  Several of these absorption lines were previously detected by Villar-Mart\'\i n et al. (1999) in their New Technology Telescope spectrum.  We also detect SiII $\lambda$1260 and SiII $\lambda$1527 blueshifted by 3600 and 3200 km s$^{-1}$, respectively, relative to the HeII emission; their blueshifts are significantly higher than those of the other lines from absorber B, suggesting that the SiII lines may come from a separate absorber.  In figure 1 we show the LRIS spectrum with the identified absorption lines marked.  Measurements of these absorption lines are given in tables 1 and 2, for absorber A and B, respectively.  The 1$\sigma$ errors associated with the equivalent width measurements were determined from the uncertainty in the height of the underlying emission.  Velocity shifts are given relative to the centroid of the HeII emission, and equivalent widths are given in the rest frame of MRC 2025-218.  We have not measured SiIV $\lambda$1394 from absorber A or SiIV $\lambda$1402.8 from absorber B, because, due to their relative velocity shifts, these two features occur at very similar wavelengths.  In the case of absorber A, we detect several excited fine structure SiII* absorption lines, and for this reason we assume that the CII absorption feature contains both CII $\lambda$1334.5 and the excited fine structure line CII* $\lambda$1335.7.  Also included in tables 1 and 2 are several lines that are undetected in absorption, but whose upper limits are useful for constraining the column densities.  

Some absorption features, although detected at the $\ge3\sigma$ level of significance in the LRIS spectrum, have not been measured.  In some cases the line identifications are ambiguous, i.e., when a line from absorber A and a line from absorber B are expected to occur at very similar wavelengths.  Several other absorption lines cannot be identified with absorber A or B, and may be formed in intervening systems at significantly lower z.  These lines are not of primary interest in this study, and shall notbe considered further.  

At the spectral resolution of the LRIS data, most of the absorption lines are unresolved in velocity space, i.e., the full width at half minimum is in agreement with the FWHM of the instrumental profile ($\sim$600 km s$^{-1}$), within errors.  For this reason, we consider that the FWHM of the absorber is likely to be $\la$600 km s$^{-1}$.  

The observed depth of the absorption line can be used to place a lower limit on the covering factor C, using the relation

\begin{equation}
C \ge \frac{F_{0}-F_{min}}{F_{0}}
\end{equation}

\noindent where F$_{0}$ is the unabsorbed flux density near the wavelength of the absorption line, and F$_{min}$ is the flux density measured at the deepest point of the absorption line profile.  For absorber A, the deepest absorption lines (CII and CIV) require C$\ge$0.7.  In the case of absorber B, the deepest of the absorption lines, SiIV $\lambda$1394 and SiII $\lambda$1527, require C$\ge$0.2.  Since the absorption lines are not well resolved spectrally, we use the curve of growth (CoG hereinafter) method to derive column densities, or limits thereto (Spitzer 1978): when an absorption line is on the linear part of the CoG, the column density N can be derived using

\begin{equation}
N = \frac{W_{\lambda} m_e c^2}{\pi e^2 f \lambda{_0}{^2}}
\end{equation}

\noindent where f is the oscillator strength and $\lambda_0$ is the rest wavelength of the line; when an absorption line is saturated, equation 2 becomes

\begin{equation}
N \ge \frac{W_{\lambda} m_e c^2}{\pi e^2 f \lambda{_0}{^2}}
\end{equation}

\noindent The implied column densities are listed in column 5 of tables 1 and 2.  While the bright emission from the active nucleus facilitates the detection of these absorption lines, the fact that this emission is spatially unresolved means that we are unable to determine the spatial extend of the absorption lines discussed above (only in the case of Ly$\alpha$ has this been possible: see $\S$3.2.1).  

\subsection{VLT VIMOS spectrum}

\subsubsection{Ly$\alpha$ absorption}
The sharp edge to the blue wing in Ly$\alpha$, and also the fact that the flux drops below the continuum level, led Villar-Mart\'\i n et al. (1999) to suggest that this line is partially absorbed by neutral Hydrogen.  Our higher spectral resolution VIMOS spectrum confirms this.  We detect two Ly$\alpha$ absorption features on the blue side of the emission line (Figures 3 and 4) with velocity shifts of $\sim$-700 and -1700 km s$^{-1}$.  We identify these two absorption features with absorber A and B, respectively.  

The absorption profile of Absorber A takes the form of a sharp absorption edge at 4402.5$\pm$0.3 \AA, with a gradual decrease in apparent optical depth towards shorter wavelengths.  At 4398.5$\pm$0.3\AA, there is secondary peak in the Ly$\alpha$ emission, which presumably marks the wavelength at which absorber A becomes relatively transparent.  At its deepest point this absorption feature is black, showing that the absorbing gas has a covering factor of 1. 

The spatial extent of the Ly$\alpha$ absorption from absorber A is strikingly large.  The sharp blue edge is detected in all fibres which have sufficient signal to noise to detect such a feature (23 fibres or 15.4 arcsec$^{2}$).  In addition, the secondary peak in the Ly$\alpha$ velocity profile is detected over a similarly large spatial extent (16 fibres or 10.7 arcsec$^{2}$).  The wavelength of the blue edge, and that of the secondary emission peak, are remarkably constant from fibre to fibre, varying by less than 100 km s$^{-1}$.  The combined spatial extent of these two features allows us to set a lower limit of 4.7$\arcsec\times3.4\arcsec$ ($\sim$40$\times$30~kpc$^{2}$) for the projected size of this Ly$\alpha$ absorber.  

Due to the complexity of the Ly$\alpha$ velocity profile, it is necessary to fit the line using model velocity profiles.  Our model profiles consist of a Gaussian shaped Ly$\alpha$ emission component, a linear continuum, and an absorption component which has a sharp absorption edge and a Voigt-like decay in column density towards bluer wavelengths.  The covering factor of the absorber, as seen by the Ly$\alpha$ nebula and the continuum, was set to 1.  Such a large covering factor is justified by the fact that the absorption feature is essentially black at the absorption edge.  All other parameters were allowed to vary freely, within physically reasonable limits (e.g., emission and absorber wavelengths: 4360-4440\AA; emission FWHM: 100-3000 km s$^{-1}$; absorber velocity width: 10-3000 km s$^{-1}$; HI column: 0-10$^{22}$ cm$^{-2}$).  We have defined the width of the absorber as the distance in velocity space from the absorption edge to where the HI column has half its peak value.  

Since we are also interested in the spatial properties of this absorption feature, we have fitted the Ly$\alpha$ profile in four apertures: (i) the `nuclear' aperture, comprising 3$\times$3 fibres (i.e. 2\arcsec$\times$2\arcsec) centred on the fibre wherein the very broad CIV emission and the continuum emission are brightest, which we assume marks the projected position of the active galactic nucleus (AGN); (ii) the `Northern' aperture, defined as 3$\times$3 fibres centred on the fibre in which Ly$\alpha$ is brightest; (iii) the `Southern' aperture, which comprises 3$\times$3 fibres with a centre that is 2 fibres South of the continuum peak; and (iv) the `spatially integrated' aperture, which is the sum of all 21 fibres contained within the above apertures.  The apertures were sized in order to maximise the signal-to-noise ratio in the line profile.  Note that the nuclear aperture is not independent of the Northern and Southern apertures.  In figure 2 we show the continuum-subtracted Ly$\alpha$ image of 2025-218, reconstructed from the VIMOS data cube (see paper for further details of this image), with extraction apertures indicated.  We have assumed that the bright continuum point-source (Pentericci et al. 1999; 2001) and the very broad CIV emission ($\sim$9000 km s$^{-1}$: Villar-Mart\'\i n et al. 1999; figure 3 of this paper) mark the position of the active galactic nucleus (AGN hereinafter).

\begin{table}
\centering
\caption{Absorber A: measurements and limits from the LRIS spectrum.  Columns: (1) atom/ion; (2) rest-frame wavelength (\AA); (3) velocity shift (km s$^{-1}$) relative to the central wavelength of the HeII $\lambda$1640 emission line; negative values indicate a blueward shift; 1$\sigma$ uncertainties are typically 100-200 km s$^{-1}$; (4) rest-frame equivalent width (\AA); (5) column density of the atom/ion.} 
\begin{tabular}{lllll}
\hline
Atom/ion & $\lambda_0$& $\Delta$v   & W$_{rest}$ & $N$ \\  
           & (\AA)    & (km s$^{-1}$)& (\AA)     & (10$^{14}$ cm$^{-2}$) \\
(1)        & (2)    & (3)  & (4) & (5) \\
\hline
CI         & 1157.2 & -670 & 2.4$\pm$0.8 & $\ge$2.5 \\
CI         & 1193.5 & -880 & 1.9$\pm$1.5 & $\ge$5.5 \\
CI         & 1277.5 & -790 & 0.6$\pm$0.3 & $\ge$1.9 \\
CI         & 1561.1 &      & $\le$0.4    & $\le$3.0 \\
CI         & 1657.2 &      & $\le$0.4    & $\le$1.2 \\
CII,CII*   & 1335.3 & -650 & 2.0$\pm$0.2 & $\ge$8.9 \\
CIV        & 1549.5 & -560 & 4.7$\pm$0.9 & $\ge$6.3 \\

NV         & 1238.8 & -460 & 0.8$\pm$0.3 & $\ge$2.2 \\

OI         & 1302.2 & -500 & 0.8$\pm$0.3 & $\ge$5.7 \\

SiI        & 1845.5 &      & $\le$0.3    & $\le$0.37 \\
SiII       & 1260.4 & -620 & 0.8$\pm$0.2 & $\ge$0.43 \\
SiII*      & 1264.7 & -580 & 0.3$\pm$0.2 & $\ge$0.08 \\
SiII*      & 1309.3 & -640 & 0.5$\pm$0.2 & $\ge$2.1 \\
SiII       & 1526.7 & -590 & 0.7$\pm$0.1 & $\ge$2.2 \\
SiII       & 1808.0 &      & $\le$0.15   & $\le$24 \\
SiII*      & 1817.2 &      & $\le$0.36   & $\le$54 \\
SiIV       & 1402.8 & -550 & 1.2$\pm$0.2 & $\ge$2.3 \\

AlII       & 1670.8 & -630 & 0.8$\pm$0.1 & $\ge$0.16 \\
AlIII      & 1854.7 & -550 & 0.7$\pm$0.1 & 0.47$\pm$0.06 \\
AlIII      & 1862.8 & -670 & 0.3$\pm$0.1 & 0.35$\pm$0.12 \\
\hline
\end{tabular}
\end{table}

\begin{table}
\centering
\caption{Absorber B: measurements and limits from the LRIS spectrum.  Columns: (1) atom/ion; (2) rest-frame wavelength (\AA); (3) velocity shift (km s$^{-1}$) relative to the central wavelength of the HeII $\lambda$1640 emission line; negative values indicate a blueward shift; 1$\sigma$ uncertainties are typically 100-200 km s$^{-1}$; (4) rest-frame equivalent width (\AA); (5) column density of the atom/ion.  The large velocity shift between SiII and the other lines suggests that the SiII lines may come from a separate absorber.} 
\begin{tabular}{lllll}
\hline
Atom/ion & $\lambda_0$& $\Delta$v   & W$_{rest}$ & $N$ \\  
           & (\AA)    & (km s$^{-1}$)& (\AA)     & (10$^{14}$ cm$^{-2}$) \\
(1)        & (2)    & (3)  & (4) & (5) \\
\hline
CI         & 1157.2 & -2500 & 1.6$\pm$0.7 & $\ge$1.4 \\
CI         & 1277.5 & -2200 & 0.6$\pm$0.2 & $\ge$2.5 \\
CII        & 1334.5 & -2200 & 0.5$\pm$0.2 & $\ge$1.5 \\

SiI        & 1845.5 &       & $\le$0.36   & $\le$0.65 \\
SiII       & 1260.4 & -3600 & 0.7$\pm$0.3 & $\ge$0.3 \\
SiII       & 1526.7 & -3200 & 0.46$\pm$0.05 & $\ge$1.5 \\
SiIV       & 1393.8 & -2500 & 0.8$\pm$0.1 & $\ge$0.80 \\
\hline
\end{tabular}
\end{table}

The fits to Ly$\alpha$ are shown in Fig 3, and the best-fitting parameters are detailed in Table 3.  However, we must emphasise that these fits are not unique: since the absorption feature is saturated (i.e., black) at some velocities, the true HI column densities could be substantially higher than these values.  Therefore, we adopt a lower limit of 10$^{14.8}$ cm$^{-2}$ for the HI column of absorber A.  This takes into account the formal uncertainty from the fitting, and represents the minimum possible HI column density for the absorber.  In addition, the fact that the Ly$\alpha$ absorption feature is saturated means the HI column density could vary significantly between apertures with no obvious difference in the observed absorption feature.  In our best fit model, the absorption edge is blueshifted by 380 km s$^{-1}$ relative to the HeII emission.  The FWHM of the unabsorbed Ly$\alpha$ profiles (1200-1600 km s$^{-1}$), are relatively large in comparison with that of HeII ($\sim$500-1100 km s$^{-1}$).  For the velocity width of the absorber itself, we obtain values in the range 250-350 km s$^{-1}$.  

Ly$\alpha$ absorption from absorber B is also detected (Figures 3 and 5), but it is not clear whether it is spatially extended.  Given the depth of this absorption feature, we estimate its covering factor C to be $\ga$0.7.  It has an equivalent width of 1.9$\pm$0.3\AA, implying a HI column density of $\ge$3.5$\times$10$^{14}$ cm$^{-2}$.  

Absorbers A appears to be kinematically distinct from the line emitting gas that comprises the EELR.  While the line emitting gas shows a clear velocity gradient across the Ly$\alpha$ nebula, the velocities at which absorber A occurs does not show any significant spatial variation.  In addition, the fact that absorbers A and B both have C$\sim$1 implies that the absorbers are very probably outside of the Ly$\alpha$ emitting region.  Since Ly$\alpha$ emission is detected at distances of up to 25 kpc from the AGN (Villar-Mart\'\i n et al. 2007), we will assume that absorbers A and B are at distances of $\ge$25 kpc from the AGN.  

\subsubsection{CIV absorption}
The CIV velocity profile (see Figs 4 and 5) also shows a sharp blue edge, which occurs at 5527 \AA, and which we identify as CIV $\lambda$1550.8 from absorber A.  This absorption feature is not black at its deepest point, and its depth implies $\tau$$\sim$1 and a CIV column of $\la$3$\times$10$^{15}$ cm$^{-2}$.  At $\sim$5618 \AA~ the velocity profile shows a peak, and blueward of this there is a second absorption feature, wherein the flux reached zero.  We identify this feature as CIV $\lambda$1548.2 absorption by absorber A.  The relative wavelengths of these two absorption features are in good agreement with the theoretical separation of the CIV doublet.  The apparent depth of the CIV $\lambda$1548.2 absorption line requires $\tau$$\ga$2 and a CIV column of $\ga$3$\times$10$^{14}$ cm$^{-2}$.  Thus, independently of model profile fitting, we can state that the column density of CIV is in the range 10$^{14.5}$-10$^{15.5}$ cm$^{-2}$.  

The velocity profile of CIV shows an additional peak blueshifted by $\sim$1500 km s$^{-1}$ relative to HeII.  Blueward of this peak, at $\lambda=$5553-5587 \AA~ (corresponding to $\sim$-1500 to -3600 km s$^{-1}$ relative to HeII), the flux dips below the continuum level.  This suggests that some of the flux within this wavelength range has been absorbed, possibly by CIV associated with absorber B.  Unfortunately, the relatively low signal to noise ratio, along with the presence of a strong sky-subtraction residual at $\sim$5577 \AA, precludes a detailed analysis of the possible CIV absorption from absorber B.  The signal to noise of our data is insufficient to determine whether the CIV absorption features are spatially extended.  

To place further constraints on the CIV column of absorber A, we have fitted model velocity profiles to the observed profile of the CIV doublet.  We restrict this analysis to the nuclear aperture, where the signal to noise is relatively high and where the CIV absorption feature is clearly detected.  In our fits, we have used two Gaussians for the CIV emission: one to represent the narrow CIV emission doublet from the EELR, and the other one to represent the CIV emission from the BLR\footnote{Treating the emission components as single lines instead of doublets has no significant impact on our model velocity profiles, because the FWHM of both emission components are substantially larger than the theoretical doublet separation.} that is very clearly seen in our VIMOS spectrum (Fig 3: FWHM$\sim$9000 km s$^{-1}$; see also the spectrum of Villar-Mart\'\i n et al. 1999).  The FWHM of the EELR emission was set to the FWHM we have measured for HeII in the nuclear aperture (1100 km s$^{-1}$).  The other parameters of the line emission components were allowed to vary freely.  

We fitted the absorption component of absorber A as a doublet, i.e. CIV $\lambda$1548 and CIV $\lambda$1551, using a pair of profiles with a sharp absorption edge and a Voigt-like decay towards shorter wavelengths.  The two absorption profiles were forced to have the same velocity shift, velocity width and N$_{CIV}$.  In order to reduce the number of free parameters, the velocity width of the CIV absorber was fixed to the value obtained from the fit to Ly$\alpha$ in the nuclear aperture.  N$_{CIV}$ and the velocity shift were allowed to vary freely.  We set the weighting of the data between 5553 \AA~ and 5587 \AA~to zero, because this region contains a strong sky subtraction residual, as well as possible CIV absorption from absorber B and possible SiII* $\lambda$1533 absorption from A.  

The best fit to the CIV velocity profile is shown in Fig 4.  Table 4 lists the parameters of the best fitting model.  We find that N$_{CIV}$=10$^{15.3\pm0.3}$ cm$^{-2}$.  The velocity shift of the CIV $\lambda$1551 absorption edge (-150 km s$^{-1}$) is in agreement with that of the Ly$\alpha$ edge, after taking into account the $\sim$200 km s$^{-1}$ uncertainty in the relative wavelength calibration between the Ly$\alpha$ and CIV regions of the spectrum.  Due to the complexity of the observed CIV velocity profile, our best fit parameters for the {\it emission} components may not be unique.  However, after repeating the fits with the FWHM and wavelength of the emission components set to a variety of different values, we found that our best fit parameters to the {\it absorption} doublet are not significantly affected by the precise profile of the CIV emission: this is due to the high optical depth of the CIV absorption lines.

\begin{figure}
\includegraphics{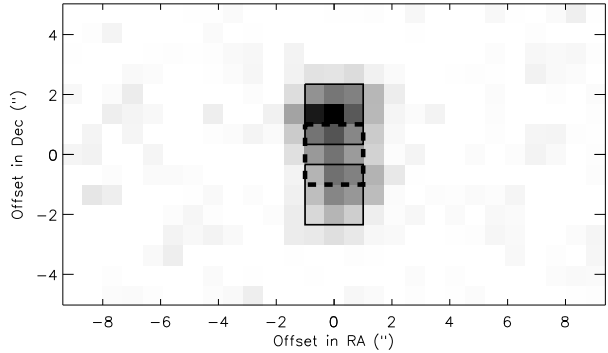}
\vspace{2.05in}
\caption{Continuum-subtracted Ly$\alpha$ image of MRC 2025-218, reconstructed from the VIMOS data cube (Paper I).  The nuclear, northern and southern apertures (see text) are marked with boxes.  North is toward the top, East is to the left.  The spatial zero of the axes corresponds to the position of the active nucleus.}
\end{figure}

\begin{figure*}
\includegraphics{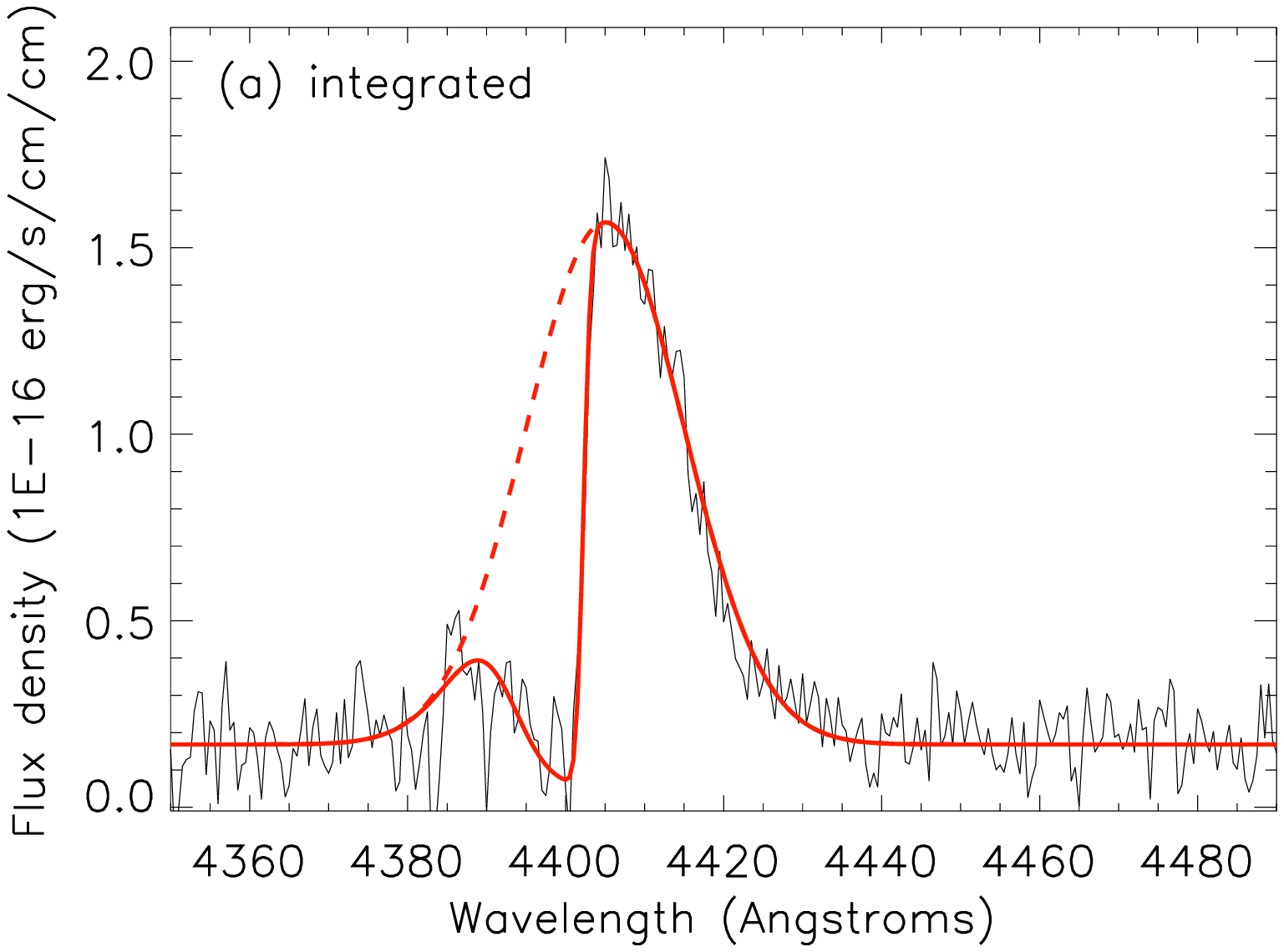}
\includegraphics{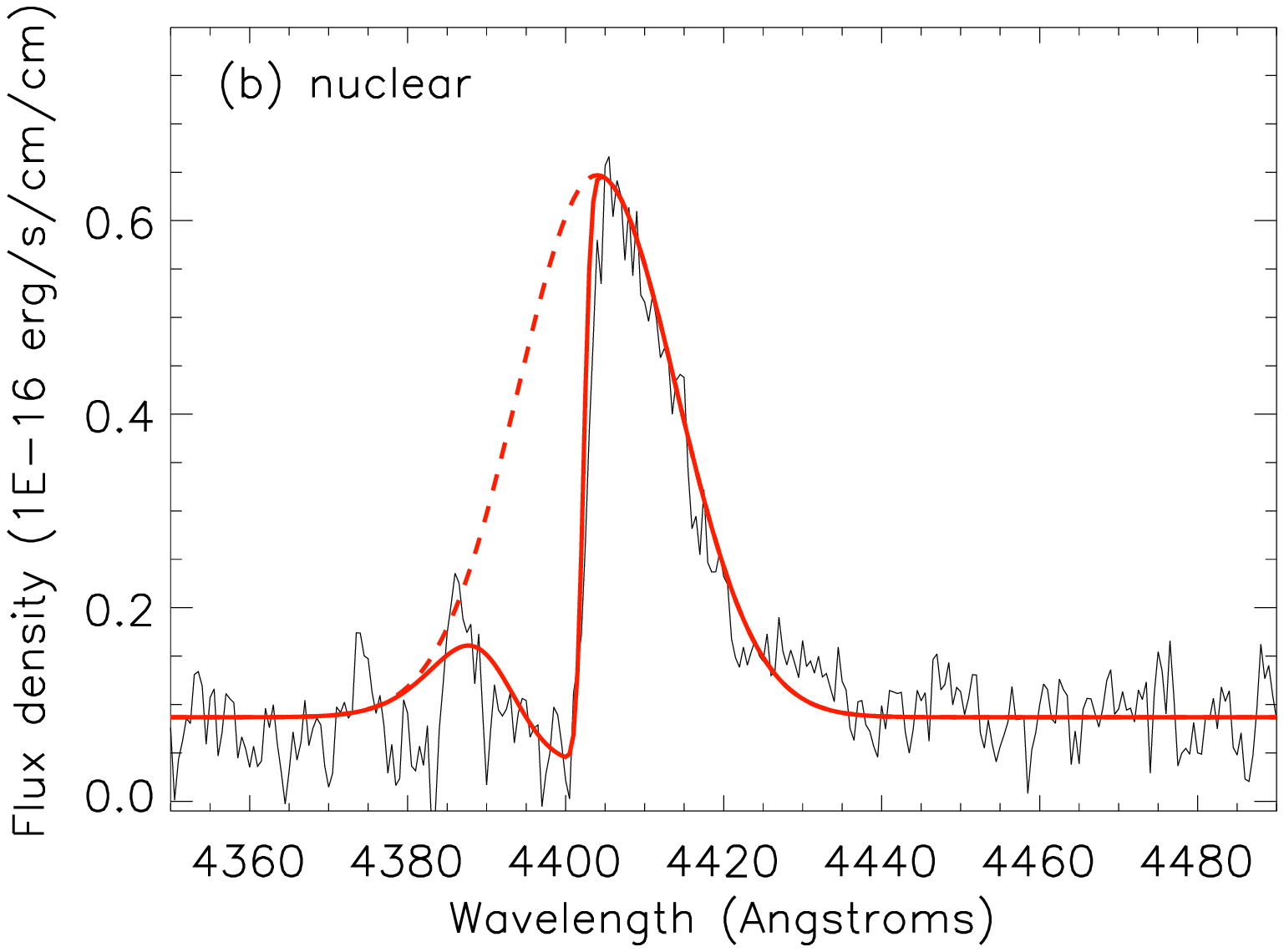}
\includegraphics{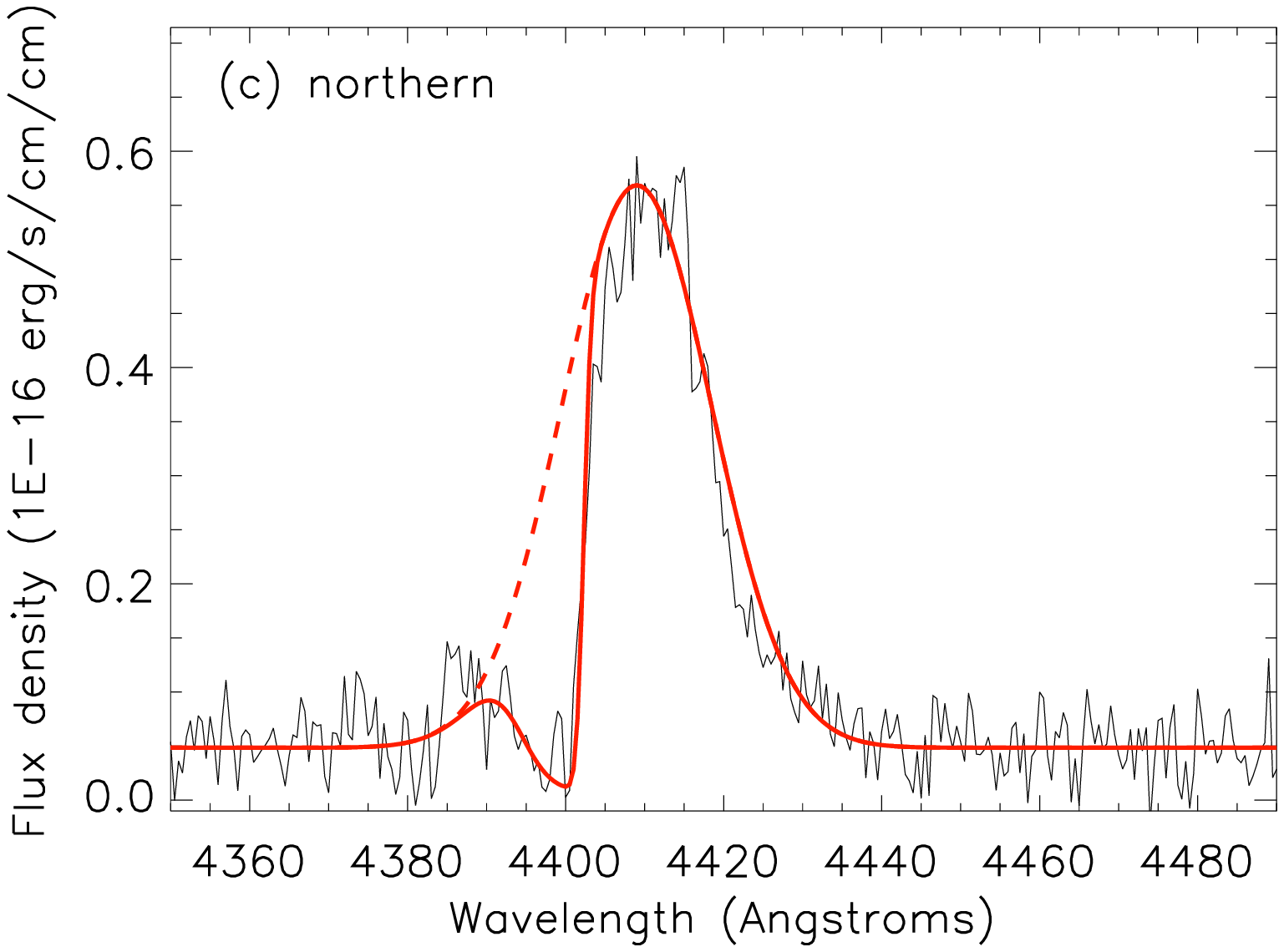}
\includegraphics{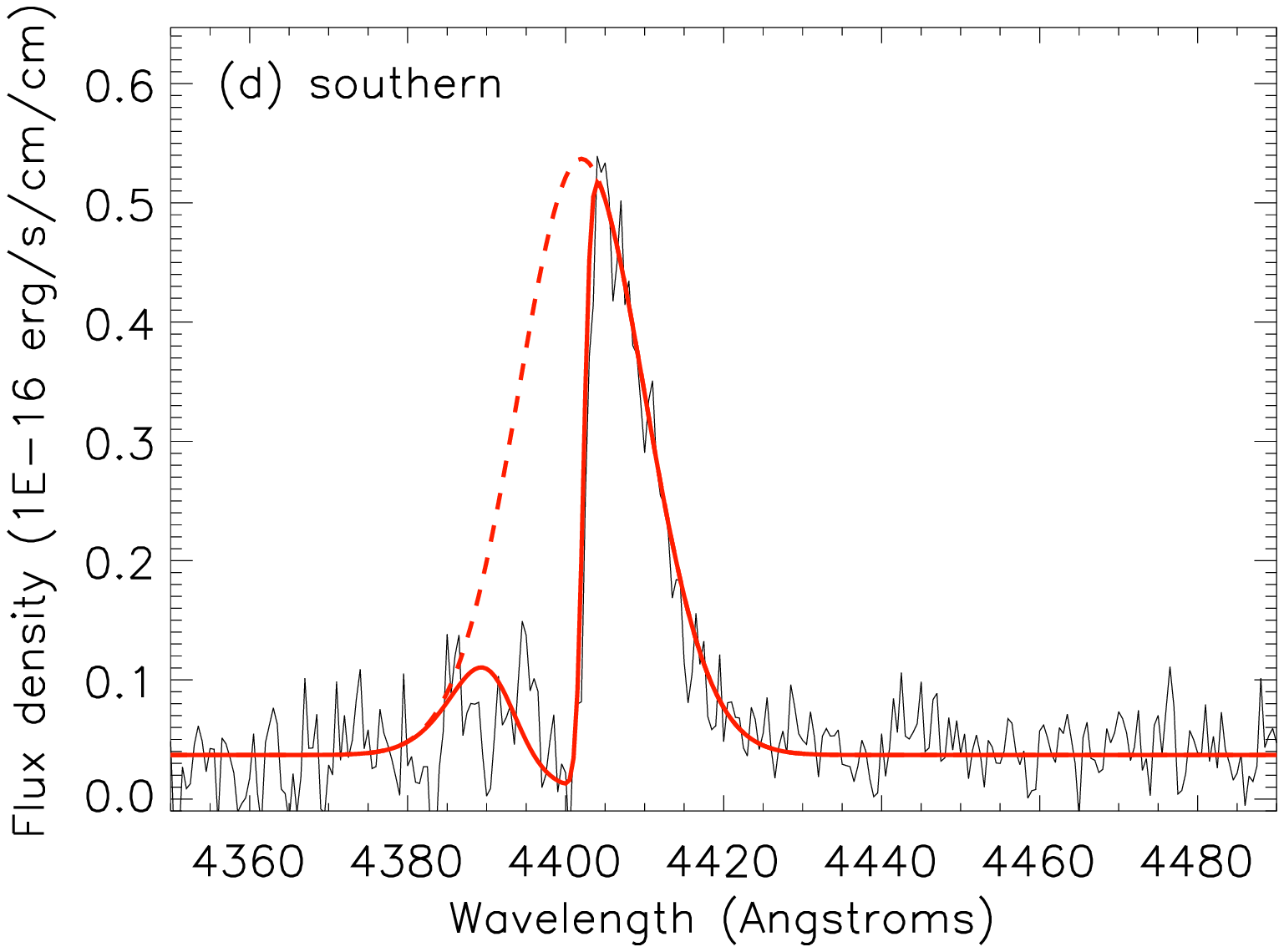}
\includegraphics{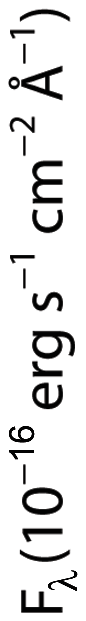}
\includegraphics{axis.eps}
\includegraphics{axis.eps}
\includegraphics{axis.eps}
\vspace{5.0in}
\caption{Fits to the Ly$\alpha$ profile of MRC 2025-218: (a) velocity profile in the spatially integrated aperture; (b) velocity profile in the nuclear aperture; (c) velocity Northern aperture; (d) velocity profile in the Southern aperture.}
\end{figure*}

\subsubsection{Other absorption lines in the VIMOS spectrum}
In addition to Ly$\alpha$ and CIV, we have detected several other absorption lines in our VIMOS spectra of MRC 2025-218.  From absorber A, we detect absorption due to NV $\lambda$1239, SiII $\lambda$1260, CII+CII* $\lambda$1335, SiIV $\lambda$1403 and SiII $\lambda$1526.  From absorber B we detect NV $\lambda$1239, CI $\lambda$1278 and SiIV $\lambda$1394.  We also detect SiII $\lambda$1260 and SiII $\lambda$1526 absorption blushifted by 3600 and 3200 km s$^{-1}$, respectively, relative to the HeII emissIn a recent paper (Villar-Mart\'\i n et al. 2007: Paper I) we reported the detection of a strong, spatially extended Ly$\alpha$ absorption feature in the spectrum of MRC 2025-218.  Absorption from SiII, SiIV, OI, CII and CIV have also been detected in the spectrum of this source (Villar-Mart\'\i n et al. 1999).  In this paper we present a more detailed analysis of the absorbing gas in front of MRC 2025-218, making use of integral field spectroscopy from the Very Large Telecope (Paper I) and a deep, long slit spectrum from the Keck II telescope (Humphrey et al. 2008).  We also examine whether the absorbing gas might be related to an outflow of gas from the host galaxy, or the cosmological infall of gas onto the host galaxy.  Throughout this paper, we assume a flat universe with $H_{0}$=71 km $s^{-1}$ Mpc$^{-1}$, $\Omega_{\Lambda}$=0.73 and $\Omega_{m}$=0.27.  At the redshift of this source, 1\arcsec corresponds to 8.1 kpc.  

ion.  As discussed earlier, it is not clear whether these lines are from absorber B or from a separate one.  In Figures 5 and 6 we show two spectral regions of the VIMOS spectrum, with absorption line identifications marked, along with corresponding regions of the LRIS spectrum for comparison.  (We do not show the 4650-5390 \AA~region of the VIMOS data because its wavelength calibration is unreliable: see $\S$2.2.)  We simply measure the equivalent width of each absorption line, and obtain column densities, or limits thereto, using the CoG method ($\S$3.1).  Tables 5 and 6 list the absorption line properties and the implied column densities.  As we discussed in $\S$3.1, we are unable to determine the spatial extent of these absorption lines because the background source, the active nucleus, is not spatialy resolved.  

\subsection{Final column densities of A and B}

For some species we have several measurements, or limits, of column density; where this is the case, we have reduced multiple measurements and limits to a `final' column density as follows.  (Unless otherwise stated, the procedures below were applied to both absorber A and absorber B.)

\begin{itemize}
\item For both NV and SiII, the possible range in column density is constrained by lower and upper limits.  In the case of NV, our adopted column density corresponds to the midpoint between the upper and the lower limit.  For SiII, we use the midpoint between the most stringent upper limit and the most stringent lower limit.  In both cases, we adopt a 1$\sigma$ uncertainty of half the difference between the two limits.  

\item In the cases of CI and SiII*, the most stringent upper and lower limits are in fair agreement (i.e., they differ by less than a factor of 2), though the upper limit is actually below the lower limit.  As before, we adopt the midpoint between these two limits, and take the 1$\sigma$ uncertainty to be half the difference between the two limits.  

\item For species with either multiple upper limits or multiple lower limits, we use the most stringent of these limits.  

\item In the case of AlIII, we simply take the mean of our two column density measurements.  

\end{itemize}

In principle, comparing the column densities of the metallic species against that of HI can provide constraints on the metallicity of the absorbing gas.  We assume that the column density of neutral Carbon relative to that of neutral Hydrogen is roughly equal to the Carbon to Hydrogen abundance ratio, i.e., $N_{CI} / N_{HI} \sim N_{C} / N_{H}$.  In the case of absorber A, the value of $N_{CI}$ and the $N_{HI}$ value from our fits to the Ly$\alpha$ profile imply a C/H abundance ratio that is $\la$600 times the solar ratio of Anders \& Grevesse (1989).  Clearly, C/H abundance ratios of several hundred times the solar value would be unrealistic in the context of gas at a distance of several tens of kpc from the galaxy nucleus; as we suggested in $\S$3.2.1.  

Conversely, by adopting a value for the C/H abundance ratio, we can then use our measurement of $N_{CI}$ to refine our lower limit on $N_{HI}$ from $\S$3.2.1.  We assume conservatively that this abundance ratio is less than ten times its solar value.  We believe this assumption to be physically reasonable in the context of spatially very extended gas outside of the galaxy nucleus.  As before, we also assume that $N_{CI} / N_{HI} \sim N_{C} / N_{H}$.  The measured CI column then implies $N_{HI} \ga 3 \times 10^{16}$.  In the case of absorber B, we only have a lower limit to $N_{CI}$, thus we cannot apply this method.  

We can apply a similar method to obtain a lower limit for the column density of ionized Hydrogen, using our measurement of $N_{SIV}$.  It is reasonable to assume that in a region of warm ionized gas, the ionization fraction SiIV/Si does not exceed the Hydrogen ionization fraction HII/H, i.e.,  $N_{SiIV} / N_{Si} \leq N_{HII} / N_{H}$.  If we also assume that $N_{Si} / N_{H} \le$10 times its solar value, the SiIV column density measurements then imply $N_{HII} \ge 7.3 \times 10^{17}$ and $\ge 2.6 \times 10^{17}$ cm$^{-2}$ for absorber A and B, respectively.  

We can also set a lower limit to the {\it total} column of Hydrogen.  For a given metal, we sum the column densities of the different ionization stages, and we then calculate the Hydrogen column density that is implied by the summed column density.  As before, we assume the metal abundances are no higher than 10 times their solar values.  In the case of absorber A, the columns of Silicon and Aluminium both imply $N_{H} \ga 2 \times 10^{18}$ cm$^{-2}$, which we adopt as our final value of the total Hydrogen column of this absorber.  (The summed column of Carbon implies $N_{H} \ga 3 \times 10^{17}$ cm$^{-2}$.)  For absorber B, the summed Silicon column implies $N_{H} \ga 7 \times 10^{18}$ cm$^{-2}$.  In Table 7 we list our final column densities.

\subsection{Mass and density}

In order to obtain crude estimates for the masses absorbers A and B, we assume that both absorbers are spherically symmetric shells, and hence $M_H = 4 \pi r^2 m_H N_H$.  Also assuming that absorber A is at a distance of $\ge$25 kpc from the AGN ($\S$3.2.1), and adopting $N_H \ga 2 \times 10^{18}$ cm$^{-2}$ ($\S$3.3), we estimate its mass to be $\ge 10^8 M_{\odot}$.  In the case of absorber B, we also adopt a distance of $\ge$25 kpc from the AGN, and hence we obtain a mass of $\ge 4 \times 10^8 M_{\odot}$.  However, a different set of assumptions could result in substantially different masses.

The ratio of the SiII and SiII* column densities (i.e., the population ratio between the $3s^{2}3p^{2}P_{1/2}$ and $3s^{2}3p^{2}P_{3/2}$ levels of SiII) is sensitive to electron density n$_{e}$, and hence can be used to obtain an estimate of the density of the absorbing gas.  For absorber A, the SiII*/SiII column ratio is $\ge$0.04 (see table 7).  Comparing this value against calculations of column ratio as a function of n$_{e}$ (Keenan et al. 1985), we find that n$_{e}\ga$10 cm$^{-3}$.  We will assume that the number density of Hydrogen $n_H$ has a similar value to n$_{e}$.  In the case of absorber B, our upper limit to the SiII*/SiII column ratio implies n$_{e}\le$1000 cm$^{-3}$.  Although excitation by 35$\mu$m IR photons can also populate the $3s^{2}3p^{2}P_{3/2}$ excited level, it seems unlikely that there would be enough IR radiation to populate the excited level in this way.  For instance, if the IR radiation from the AGN and host galaxy were responsible for populating the $3s^{2}3p^{2}P_{3/2}$ SiII* level, then a specific luminosity at 35$\mu$m of $\sim 10^{39}$ erg s$^{-1}$ Hz$^{-1}$ would be required.  However, the specific IR luminosities of powerful active galaxies at z$\sim$2-3 are more than 4 orders of magnitude lower than this (e.g. Seymour et al. 2007).  

\subsection{Ionization of absorber A}

Several previous studies have considered the possible source of ionization of the absorbers around high-z radio galaxies.  Binette et al. (2006) performed detailed photoionization calculations for the spatially extended absorbers around two other high-z radio galaxies, with the goal of explaining the measured $N_{CIV}/N_{HI}$ ratio.  They found that hot stars, or a metagalactic background radiation (MBR hereinafter) field in which stars dominate over AGN, are equally successful for reproducing this column ratio.  We now consider various sources of ionization for absorber A.  

\noindent {\bf Young stars.}  The measured $N_{NV}/N_{CIV}$ ratio of 1.8$_{-1.2}^{+3.2}$ requires a hard ionizing continuum, and hence we reject young stars as the main source of ionization.  

\noindent {\bf Metagalactic background.}  Adopting the intensity $J_v = 10^{-21}$ erg cm$^{-2}$ s$^{-1}$ Hz$^{-1}$ sr$^{-1}$ inferred by Cooke et al. (1997), and using the energy distribution of Fardal, Giroux \& Schull (1998), we estimate that the MBR has an ionizing photon flux of $\sim 2 \times 10^6$ cm$^{-2}$ s$^{-1}$.  This is roughly in agreement with the flux needed to explain the HII column of the absorber ($\alpha_{rec} n_e N_{HII} \ga 10^6$ cm$^{-2}$ s$^{-1}$).  Thus, the ionization of absorber A can be explained with photoionization by the MBR. 

\noindent {\bf Active nucleus.}  The clear detection of an unresolved continuum source in rest-frame UV and optical images of MRC 2025-218 (e.g. Pentericci et al. 1999; 2001; also Paper I) indicates that the active nucleus is not completely obscured along our line of sight.  Might the radiation field of the active nucleus be responsible for the ionization of absorber A?  We extrapolate the ionizing photon flux from the observed continuum flux at 1260 \AA, adopting the spectral energy distribution of Korista et al. (1997), with the exponential cut off set to $kT=$86 eV, and assuming that the continuum is reddened with A$_v \ge$1.4 mag (Larkin et al. 2000).  We estimate the ionizing photon flux r=25 kpc from the active nucleus to be $\le 10^8$ s$^{-1}$ cm$^{-2}$, compared to the flux of 10$^6$ cm$^{-2}$ s$^{-1}$ required to explain the HII column of absorber A.  Thus, the radiation field of the active nucleus could also be responsible for the ionization of absorber A.  

\noindent {\bf Shocks.}  In explaining the ionized gas associated with active galaxies, shocks are often considered as an alternative to photoionization (e.g. Dopita \& Sutherland 1995).  Here, we compare the column density ratios measured in absorber A against those predicted by the fast shock models of Dopita \& Sutherland (1996: DS96 hereinafter).  We find that the pure shock models (i.e., ignoring the photoionized precursor region) are unable to explain the column ratios measured in absorber A: these models predict a much lower ionization state than is implied by our measurements, i.e., in absorber A we measure $N_{SiIV}$/$N_{SiII} \ge 0.1$, but in the DS96 models the $N_{SiIV}$/$N_{SiII}$ ratio is at least 100 times lower.

\noindent {\bf Shock continuum.}  Although pure shocks are unable to explain the column density ratios of absorber A, it is interesting to consider whether the strong UV radiation field emitted by hot, post shock cooling plasma (e.g. Daltabuit \& Cox 1972; Dopita, Binette \& Schwartz 1982) might be able to explain the ionization of absorber A.  To explain the HII column of the absorber, the shock would need to have a surface area of $\ga 1.5 \times 10^{43}$ cm$^2$ (or $\ga 1.6$ kpc$^2$), and contain a gas mass of $\ga 10^7 M_{\odot}$ (see DS96).  To put this into perspective, we estimate that the EELR contains $\sim10^9$ M$_{\odot}$ of warm ionized gas, based on the Ly$\alpha$ lumnosity of the EELR (e.g. Villar-Mart\'\i n et al. 2003).  Is photoionization by shocked gas able to explain the column ratios of absorber A?  To explore this, we have calculated photoionization models using the multipurpose code $\tt MAPPINGS$ Ic (Binette, Dopita \& Tuohy 1985; Ferruit et al. 1997), consisting of an isobaric, plane parallel slab of gas onto which an ionizing continuum impinges.  We adopt a spectral energy distribution of $F_v \propto v^{-1.5}$ as an approximation of the ionizing continuum emitted by the shocked gas [see Fig 3 of DS96; our results would not be significantly affected if we use a slightly softer ($F_v \propto v^{-2.0}$) or slightly harder ($F_v \propto v^{-1.0}$) energy distribution].  In order to satisfy simultaneously the observed column ratios $N_{CIV}/N_{CII} \le 0.4$ and $N_{SiIV}/N_{SiII} \ge 0.11$, an ionization parameter of $U=Q/(4 \pi d^2 n_H c)=0.0005-0.005$ would be required, implying $N_{HII}=7.5 \times 10^{19}$ to $7.5 \times 10^{19}$ cm$^{-2}$ and an ionizing flux of $5 \times 10^{9}$ to $5 \times 10^{10}$ cm$^{-2}$ s$^{-1}$, consistent with the lower limits derived earlier in this paper.  

In summary, there are three plausible explanations for the ionization of absorber A: the metagalactic background radiation; the radiation field of the active nucleus; or continuum emitted by shocked gas.

\begin{table*}
\centering
\caption{Results from fitting the Ly$\alpha$ profile.  Columns: (1) name of aperture; (2) central wavelength of emission Gaussian (\AA); (3) FWHM of emission Gaussian (km s$^{-1}$); (4) wavelength of the peak column density of the absorption component (\AA); (5) column density of the absorption component in logarithm (cm$^{-2}$); (6) Velocity width of the absorption component, defined as the distance in velocity space from the absorption edge to the velocity where the HI column reaches half its peak value (km s$^{-1}$); (7) Velocity shift, relative to the central wavelength of the HeII $\lambda$1640 emission line (km s$^{-1}$), of the peak optical depth of the Ly$\alpha$ absorber; the 1$\sigma$ errors take into account fitting uncertainties and the uncertainty in the wavelength calibration; the fitting uncertainties alone are 20-30 km s$^{-1}$.} 
\begin{tabular}{llllllll}
\hline
Aperture   & $\lambda_{em}$ & FWHM$_{em}$ & $\lambda_{ab}$ & Log N$_{HI}$ & Width & $\Delta$v$_{ab}$ \\  
(1)        & (2)        & (3)          & (4) & (5)            & (6)          & (7)  \\
\hline
Integrated & 4405$\pm$1 & 1600$\pm$200 & 4402.1$\pm$0.3 & 15.1$\pm$0.3 & 300$\pm$60 & -380$\pm$160 \\
Nucleus    & 4404$\pm$1 & 1600$\pm$200 & 4402.2$\pm$0.3 & 15.1$\pm$0.3 & 350$\pm$60 & -380$\pm$160 \\
North      & 4409$\pm$1 & 1500$\pm$200 & 4402.1$\pm$0.4 & 15.1$\pm$0.3 & 250$\pm$60 & -380$\pm$160 \\
South      & 4402$\pm$1 & 1200$\pm$200 & 4402.0$\pm$0.3 & 15.2$\pm$0.3 & 300$\pm$50 & -390$\pm$160 \\
\hline
\end{tabular}
\end{table*}

\begin{table*}
\centering
\caption{Results from fitting the CIV profile in the nuclear aperture.  Columns: (1) central wavelength of BLR emission Gaussian (\AA); (2) FWHM of BLR emission Gaussian (km s$^{-1}$); (3) central wavelength of EELR emission Gaussian (\AA); (4) FWHM of EELR emission Gaussian (km s$^{-1}$); (5) central wavelength of the absorption component (\AA); (6) column density of the absorption component (cm$^{-2}$) in Logarithm; (7) Velocity width of the absorption component (km s$^{-1}$; constrained from the Ly$\alpha$ fit); (8) Velocity shift, relative to the central wavelength of the HeII $\lambda$1640 emission line (km s$^{-1}$), of the peak optical depth of the CIV absorber; the uncertainty takes into account the uncertainty in the fit, and also the uncertainty in the wavelength calibration; the fitting uncertainty is 32 km s$^{-1}$.} 
\begin{tabular}{llllllll}
\hline
$\lambda_{BLR}$ & FWHM$_{BLR}$ & $\lambda_{EELR}$ & FWHM$_{EELR}$ & $\lambda_{ab}$ & Log N$_{CIV}$ & Width & $\Delta$v$_{ab}$ \\  
(1) & (2) & (3) & (4) & (5) & (6) & (7) & (8) \\
\hline
5606$\pm$5 & 9000$\pm$800 & 5614$\pm$2 & 1100 & 5615.3$\pm$0.6 & 15.3$\pm$0.2 & 350 & -150$\pm$130 \\
\hline
\end{tabular}
\end{table*}

\begin{figure}
\includegraphics{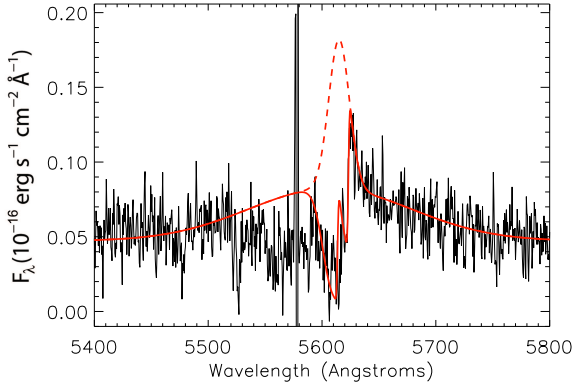}
\includegraphics{axis.eps}
\vspace{2.1in}
\caption{Fit to the CIV profile of MRC 2025-218 in the nuclear aperture.  The weighting of the data between 5520 \AA~ and 5590 \AA~was reduced, because this region is strongly affected by residuals from the sky-subtraction (this includes a large spike at $\sim$5580 \AA).}
\end{figure}

\begin{figure}
\includegraphics{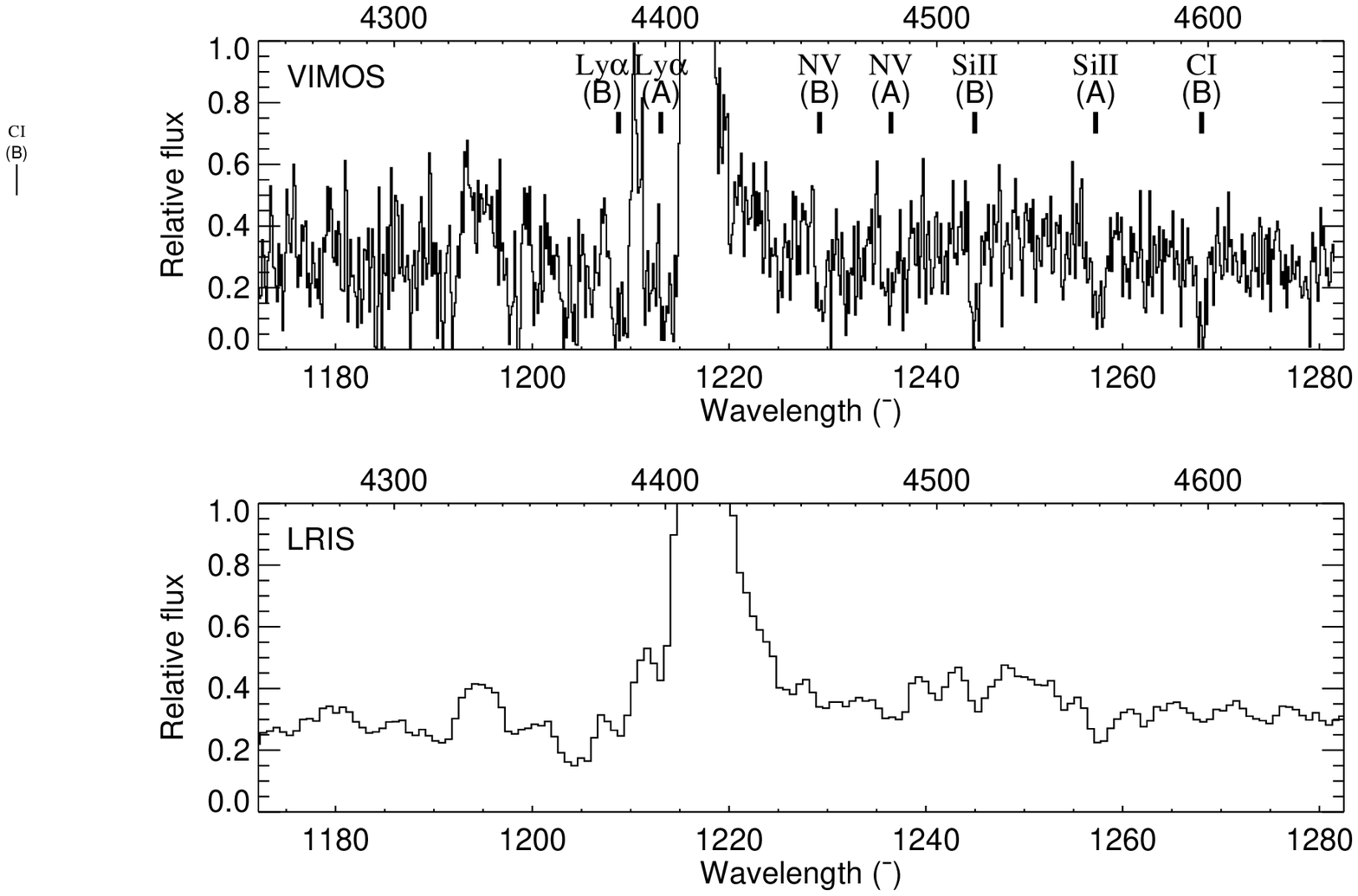}
\includegraphics{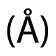}
\includegraphics{ang.eps}
\vspace{2.4in}
\caption{Absorption lines detected in the blue part of the VIMOS spectrum (upper panel); the spectrum was extracted from the fibre associated with the nucleus, to maximise the signal to noise.  For comparison, we also show the corresponding spectral region of the LRIS spectrum (lower panel).}
\end{figure}

\begin{figure}
\includegraphics{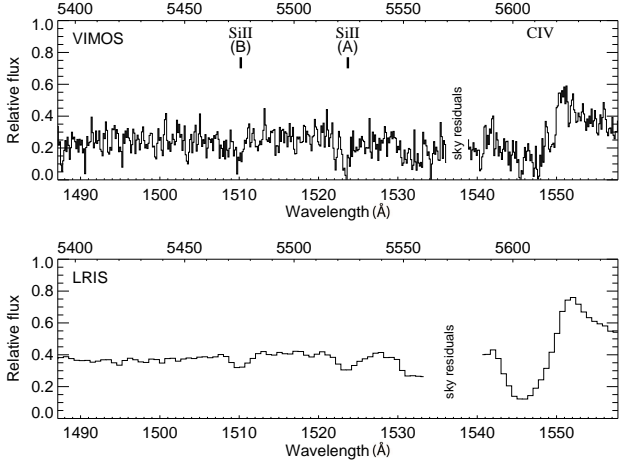}
\includegraphics{ang.eps}
\includegraphics{ang.eps}
\vspace{2.4in}
\caption{Absorption lines detected in the red part of the VIMOS spectrum (upper panel); the spectrum is from the Nuclear aperture.  For comparison, we also show the corresponding spectral region of the LRIS spectrum (lower panel).}
\end{figure}

\begin{table}
\centering
\caption{Absorber A: measurements and limits for other lines in the VIMOS spectrum.  Columns: (1) ion; (2) rest-frame wavelength (\AA); (3) velocity shift (km s$^{-1}$) relative to the central wavelength of the HeII $\lambda$1640 emission line; negative values indicate a blueward shift; 1$\sigma$ uncertainties are 100-200 km s$^{-1}$; the CII,CII* and SiIV lines fall in the region of the spectrum where the wavelength calibration is inreliable -- for these lines reason we have not calculated velocity shifts; (4) rest-frame equivalent width (\AA); (5) column density of the ion.} 
\begin{tabular}{lllll}
\hline
Ion & $\lambda_0$& $\Delta$v   & W$_{rest}$ & $N$ \\  
           & (\AA)    & (km s$^{-1}$)& (\AA)     & (10$^{14}$ cm$^{-2}$) \\
(1)        & (2)    & (3)  & (4)         & (5) \\
\hline
CII,CII*   & 1335.3 &      & 1.9$\pm$0.2 & $\ge$8.4 \\
NV         & 1238.8 & -570 & 0.8$\pm$0.4 & $\ge$2.2 \\
NV         & 1242.8 &      & $\le$0.4    & $\le$4.9 \\
SiII       & 1260.4 & -690 & 1.2$\pm$0.3 & $\ge$0.28 \\
SiII*      & 1264.7 &      & $\le$1.0    & $\le$1.2 \\
SiII       & 1526.7 & -460 & 0.7$\pm$0.1 & $\ge$2.2 \\
SiIV       & 1402.8 &      & 1.4$\pm$0.2 & $\ge$2.7 \\
\hline
\end{tabular}
\end{table}

\begin{table}
\centering
\caption{Absorber B: measurements and limits for other lines in the VIMOS spectrum.  Columns: (1) atom/ion; (2) rest-frame wavelength (\AA); (3) velocity shift (km s$^{-1}$) relative to the central wavelength of the HeII $\lambda$1640 emission line; negative values indicate a blueward shift; 1$\sigma$ uncertainties are 100-200 km s$^{-1}$; SiIV falls in the region of the spectrum where the wavelength calibration is unreliable, and for this reason we have not calculated its velocity shift; (4) rest-frame equivalent width (\AA); (5) column density of the atom/ion.  The large velocity shift between SiII and the other lines suggests that the SiII lines may come from a separate absorber.} 
\begin{tabular}{lllll}
\hline
Atom/ion & $\lambda_0$& $\Delta$v   & W$_{rest}$ & $N$ \\  
           & (\AA)    & (km s$^{-1}$)& (\AA)     & (10$^{14}$ cm$^{-2}$) \\
(1)        & (2)    & (3)  & (4) & (5) \\
\hline
HI         & 1215.7 & -1700 & 1.9$\pm$0.3 & $\ge$3.5 \\
CI         & 1277.5 & -2200 & 0.9$\pm$0.1 & $\ge$5.0 \\
NV         & 1238.8 & -2300 & 0.8$\pm$0.3 & $\ge$2.3 \\
SiII       & 1260.4 & -3700 & 1.1$\pm$0.1 & $\ge$0.71 \\
SiII*      & 1264.7 &       & $\le$0.9    & $\le$0.70 \\
SiII       & 1526.7 & -3200 & 0.6$\pm$0.3 & $\ge$1.1 \\
SiIV       & 1393.8 &       & 1.0$\pm$0.2 & $\ge$0.92 \\
\hline
\end{tabular}
\end{table}

\begin{table}
\centering
\caption{Physical properties of absorbers A and B.  The first 15 lines list the column densities, which are given in units of 10$^{14}$ cm$^{-2}$.  The final four lines give the distance of the absorber from the AGN, the thickness of the absorber, its mass and its density.} 
\begin{tabular}{lll}
\hline
Atom/ion   & A           & B \\  
\hline
HI         & $\ga$300    & $\ga$3.5 \\
HII        & $\ga$7300   & $\ga$2600 \\
H          & $\ga$20000  & $\ga$70000 \\
CI         & 1.4$\pm$0.3 & $\ge$5 \\
CII        & $\ge$9.9 &  $\ge$1.5 \\
CIV        & 2.0$^{+2.0}_{-1.0}$ & \\
C          & $\ge$12   & \\
NV         & 3.6$\pm$1.4 & $\ge$2.3 \\
OI         & $\ge$6.8    &           \\
SiI        & $\le$0.37   & $\le$0.65 \\
SiII       & 13$\pm$11   & $\ge$1.5 \\
SiII*      & 1.7$\pm$0.7 & $\le$0.71 \\
SiIV       & $\ge$2.6    & $\ge$0.92 \\
AlII       & $\ge$0.16   & \\
AlIII      & 0.4$\pm$0.1 & \\
r          & $\ge$25 kpc & $\ge$25 kpc \\
$M$        & $\ge 10^{8} M_{\odot}$ & $\ge 4 \times 10^{8} M_{\odot}$ \\
$n_H$      & $\ge$10 cm$^{-3}$ & $\le$1000 cm$^{-3}$\\
\hline
\end{tabular}
\end{table}

\section{Origin of absorber A}
Understanding the precise nature of the absorbing gas around high-z galaxies requires knowledge of the systemic velocity of the host galaxy.  Since we do not detect photospheric lines in the spectrum of MRC 2025-218, we do not know the systemic velocity of the host galaxy, and thus we are unable to distinguish whether absorber A is in infall or outflow.  Therefore, both possibilities will be considered.  As is usually the case with absorbers in front radio galaxies, we are unable to rule out that this absorber is not associated with the MRC 2025-218; however, in the interest of brevity we will not give further consideration to this particular possibility.

\subsection{Outflow}
Much evidence has accumulated in support of the idea that part of the ionized gas of some high-z radio galaxies is in outflow.  The wedge-shaped Ly$\alpha$-emitting structure oriented perpendicularly to the radio axis in TN J1338-1942 (z=4.1; Zirm et al. 2005) and the extreme emission line kinematics in the EELR of 1138-262 (z=2.2) are particularly spectacular examples of this phenomenon.  The preponderance of blue asymmetries in the emission lines provides further evidence for the outflow of ionized gas from high-z radio galaxies (Jarvis et al. 2003; Humphrey et al. 2006).  

A number of previous investigator have proposed scenarios in which the {\it absorbers} around high-z radio galaxies are also in outflow (see e.g. Krause 2002; Jarvis et al. 2003; Wilman et al. 2004).  In this section, we will examine whether it is feasible that absorber A is in outflow from the host galaxy.

The kinematics of the absorbing gas appear consistent with those of starburst driven outflows observed in absorption in star forming galaxies.  Specifically, we observe a sharp absorption edge in the blue wing of Ly$\alpha$ and CIV, where the flux goes to zero, with Ly$\alpha$ also showing a gradual decrease in the apparent optical depth towards shorter wavelengths (cf. Conti et al. 1996; Pettini et al. 2000).  

In the simplest scenario, supernova explosions associated with a massive starburst produce an expanding bubble of hot gas.  The superbubble itself is expected to contain hot gas (T$\sim 10^6$ K) which, due to its high level of ionization, would not be observable in the UV absorption lines considered herein.  At some point in its expansion, the gas at the boundary of the bubble comes into pressure equilibrium with the ambient intergalactic medium (IGM), and cools very efficiently thereafter.  This cooled gas would be detectable in various UV absorption lines.  In the case of MRC 2025-218, we suggest that absorption feature A might be due to gas that has cooled and accumulated at the interface between a superwind bubble and the ambient IGM.  

In $\S$3 we derived a lower limit to the mass of absorber A of $\ge 10^{8} M_{\odot}$.  If we were to adopt an outflow velocity of $\sim$700 km s$^{-1}$ (the typical blueshift of absorber A relative to the EELR: $\S$3), the kinetic energy of absorber A would then be $\ga 5 \times 10^{56}$ erg.  Could supernovae (SNe) inject sufficient energy into the ISM to power an outflow with this kinetic energy?  Adopting a standard initial mass function, SNe yield $\sim$10$^{49}$ erg $M_{\odot}^{-1}$ (Benson et al. 2003), meaning that a kinetic energy of $\ga 5 \times 10^{56}$ erg would thus require a stellar mass of at least $5 \times 10^{7} M{\odot}$.  This is compatible with the upper limit to the stellar mass of 4$\times$10$^{11} M_{\odot}$ for 2025-218 (Seymour et al. 2007).  Thus, if the absorber associated with 2025-218 is in outflow, it is energetically feasible that it is powered by SNe.  

The star formation rate would represent a further consistency check, because the rate at which SNe occur is also an important for powering an outflow.  Unfortunately, the star formation rate in the host galaxy of MRC 2025-218 is unknown.  Indeed, as yet there is no evidence for the presence of young stars therein (e.g. Cimatti et al. 1994).  Sub-mm observations (e.g. Reuland et al. 2004) would be useful to estimate the rate of star formation in MRC 2025-218.  

Would the radio jets of MRC 2025-218 also be able to provide enough energy to power this outflow?  Following Nesvadba et al. (2006), the radio luminosity of MRC 2025-218 (e.g. Carilli et al. 1997) implies a kinetic luminosity $L_{kin}$ on the order of $\sim$10$^{46}$ erg s$^{-1}$.  The age of the radio source can be estimated by dividing its projected radius by its advance speed.  For a projected radio source radius of 21 kpc (Carilli et al. 1997), and assuming an advance speed of $\sim$0.1c (e.g. Humphrey et al. 2007), we obtain an age of $\sim$10$^{5}$ yr.  Multiplying the kinetic energy of the radio source by its age then yields an energy of $\sim$10$^{58}$ erg.  Thus, it appears that the radio source is sufficiently energetic to have powered the gaseous outflow.  However, we are not convinced that the anisotropic radio jets would be able to produce an outflow with the observed kinematic uniformity across such a large spatial extent.  

We note that in MRC 2025-218, we do not find evidence to suggest the extended ionized gas is in outflow.  If absorber A is outflowing, it seems strange that the ionized gas at smaller radii does not show disturbed kinematics (Nesvadba et al. 2006) or morphological signatures suggestive of outflow (Zirm et al. 2005).

\subsection{Infall}
There has been much interest in recent years in the likely spectral signatures of the cosmological infall of gas onto galaxies at high-z (e.g. Haiman, Spaans \& Quartaert 2000).  Indeed, a number of recent investigations have revealed evidence for gas infall onto high-z active galaxies (Barkana \& Loeb 2003; Weidinger, M\o ller \& Fynbo 2004; Humphrey et al. 2007).  Different streams of infalling matter are expected to collide at supersonic velocities, resulting in a quasi-spherical shock wave at or beyond the virial radius (e.g. Keshet et al. 2003).  

According to the model of Barkana \& Loeb (2003), the accretion shock is expected to result in (i) an observable absorption feature in Ly$\alpha$, with a sharp edge marking the red side of this feature which corresponds to the infall velocity of Hydrogen crossing the shock, and (ii) a second Ly$\alpha$ emission peak a few hundred km s$^{-1}$ blueward of the absorption edge, due to the decreasing gas density towards larger radii and smaller infall velocities.  Both of these features are present in the Ly$\alpha$ profile of MRC 2025-218 ($\S$3.1).  We suggest that the radiation field produced by the post shock, cooling gas (e.g. DS96) might be responsible for ionizing the gas that has not yet crossed the shock (i.e., absorber A: see $\S$3.5).  

Due to the relatively large expected radius of the accretion shock ($\sim$100 kpc: Barkana \& Loeb 2003), we expect the absorption feature to have a large spatial extent and little variation in redshift across its full spatial extent.  Absorber A associated with MRC 2025-218 is consistent with the infall model of Barkana \& Loeb in this respect also.  

At the redshift of 2025-218, their model predicts a rate of gas accretion onto the host galaxy of $\sim$200 M$_{\odot}$ yr$^{-1}$.  It is interesting to note that this accretion rate is in reasonable agreement with the order of magnitude infall rates that have been calculated for the EELR of (other) radio galaxies at z$\sim$2.5 (a few hundred M$_{\odot}$ yr$^{-1}$: Humphrey et al. 2007).  Analytical infall models imply a volume averaged Hydrogen density of $\rho_H \sim$20 times the cosmic mean density, i.e., $\sim 10^{-4}$ cm$^{-3}$, at the radius of the accretion shock (Barkana 2004).  A volume filling factor of $\la 10^{-7}$ would thus be required to reconcile the measured gas number density $n_e \ge 300$ cm$^{-3}$ of absorber A with the predicted $\rho_H$.  

Determining the metallicity of the absorbing gas is important for discriminating between infall and outflow.  Naively, low metallicities would be expected if the absorbing gas is in infall (i.e., $\la$1 per cent solar: e.g. Shaye et al. 2003), while outflowing gas would be expected to have a metallicity of $\ga$10 per cent solar.  Unfortunately, in the case of MRC 2025-218 our spectra have not allowed us to place any useful contraints on the metallicity of the absorbing gas.  

Using the analytical model of Barkana (2004) for the infall of gas onto a dark matter halo, we can estimate the column of H within a particular range of line of sight velocity (LOSV).  The Ly$\alpha$ absorber in 2025-218 is detected across a range in LOSV of $\sim$270 km s$^{-1}$ (Fig 2: 4398.5 \AA~to 4402.5 \AA), and thus we will consider only the H column from the LOSV of the accretion shock to a LOSV $\sim$270 km s$^{-1}$ blueward of this.  We assume that the accretion shock radius is $\sim$100 kpc and that the infalling gas has a velocity of 550 km s$^{-1}$ when it reaches the shock (Barkana \& Loeb 2003).  Based on these assumptions, we estimate that the infall rate of 200 M$_{\odot}$ yr$^{-1}$ predicted from the model of Barkana \& Loeb would imply a H column of roughly 3$\times$10$^{19}$ cm$^{-2}$ within the relevant LOSV range.  This estimate is likely to be a lower limit, because the post-shock cooling gas has not been considered.  The H column density we have estimated for absorber A is not inconsistent with this value.  

Since the true systemic velocity is also unknown in other high-z radio galaxies, we suggest that, even when blueshifted relative to the line emission, the absorbers associated with some galaxies might in fact be in infall, rather than in outflow as is usually supposed.  Indeed, the HI columns of a number of the absorbers (e.g. Wilman et al. 2004) are in fair agreement with the HI column expected from the infall scenario of Barkana \& Loeb (2003).  By the same token, we suggest that the Ly$\alpha$ absorbers sometimes associated with non-active galaxies at high-z (e.g. LAB-2: Wilman et al. 2005) might also be related to the accretion of gas by the host galaxy.  

\section{Absorber B}
The precise nature of absorber B is not clear.  Its relatively large blushift with respect to the EELR emission of MRC 025-218 allows us to reject infall.  One possibility is that absorption component B is in outflow from MRC 2025-218.  A starburst driven super wind or a radio jet driven outlow would be energetically feasible (see $\S$4.1).  There is some evidence to suggest that MRC 2025-218 is located in galaxy group (McCarthy, Persson \& West 1992); we consider absorption arising in an intervening starburst galaxy to be a possible alternative to outflow.  The detection of NV in absorption implies that young stars cannot be the only source of ionization for absorber B.  Unfortunately, our measurements do not allow us to place any further constraints on the source of ionization.  

\section{Concluding remarks}
There is an extensive body of literature discussing absorption lines associated with active galaxies (see e.g. Hamann \& Ferland 1999; Crenshaw et al. 1999; and references therein).  In essentially all of these investigations, the source against which the absorbing gas is seen is spatially unresolved, i.e., it has a size equivalent to that of the inner accretion disk.  The overarching difficulty of these studies is to arrive at a firm conclusion about the nature of the absorber, i.e., it is not usually clear whether the absorber is a tiny plume of gas floating above the accretion disc, or a filament in the BLR located 1 pc from the accretion disc, or part of a galactic-scale super-wind, etc.  Moreover, when discussing a spatially unresolved absorber, it is not immediately clear what such a small region of gas is representative of.  

By contrast, the study of spatially extended absorbers around galaxies is a relatively new approach to obtaining information about galaxies and their environments (e.g. van Ojik et al. 1997; Binette et al. 2006; Wilman et al. 2005; Wilman, Edge \& Swinbank 2006).  Giant nebulae of Ly$\alpha$ emission (e.g. Villar-Mart\'\i n et al. 2003) can provide a relatively bright and spatially extended background source by which intervening cold and warm gas clouds can be studied (van Ojik et al. 1997; Jarvis et al. 2003). If the absorber is found to be spatially extended, then its location in relation to the galaxy nucleus can be more readily ascertained.  Unfortunately, HzRG typically have relatively faint rest-frame UV continuum emission, and this normally precludes the detection of the multitude of other strong absorption lines that are expected to be present at wavelengths where line emission is very weak or absent.  With only measurements of the Ly$\alpha$ absorption line, and sometimes also CIV, it can be difficult to obtain definitive conclusions about the physical conditions and metallicity of the absorbing gas (see Binette et al. 2006 and references therein).  We are of the opinion that galaxies that have both spatially extended Ly$\alpha$ emission and very bright continuum emission hold much potential for further elucidating the true nature of associated absorbers.  

In this paper, we have presented a study of the absorption lines in the rest-frame UV spectrum of one such galaxy: the z$=$2.63 radio galaxy MRC 2025-218.  Thanks to the bright and spatially extended Ly$\alpha$ emission, we have been able to trace Ly$\alpha$ absorption across $\sim 40\times30$ kpc$^{2}$, where it shows remarkably little spatial variation in its properties.  This absorber is kinematically detached from the extended Ly$\alpha$ emitting gas, and has a covering factor of $\sim$1.  Based on these properties, we have concluded that the absorber is outside of the EELR, i.e., at a distance of $\ga$25 kpc from the galactic nucleus.

In addition, MRC 2025-218 shows bright continuum emission from the active nucleus which has allowed us to measure metal absorption lines from a variety of atoms/ions and ionization levels.  Although the continuum source, and hence the absorption lines, are not spatially resolved, we associate them with the spatially extended Ly$\alpha$ absorber since their redshifts and velocity widths are consistent, with the formal uncertainties.  We find that the column density ratios are most naturally explained with photoionization by a hard continuum; an ionization parameter U$\sim$0.0005-0.005 is implied.  We reject shocks and also photoionization by young stars, since neither is able to reproduce satisfactorily the measured column ratios.  We have derived a lower limit of $\ge$10 cm$^{-3}$ for the electron density of the absorber.  The data do not allow useful constraints to be placed on the metallicity of the absorber.  

We have considered two possibilities for the nature of this absorber: the cosmological infall of gas, and an outflow driven by supernovae or the radio-jets.  We find good agreement between the observed properties of the HI absorber and the properties of the HI absorption expected from the cosmological infall model of Barkana \& Loeb (2003).  It is also plausible that the absorber around MRC 2025-218 is in outflow.  Discriminating between infall and outflow would require knowledge of the systemic velocity of the host galaxy, and one way to achieve this could be to measure CI and CO lines using e.g. ALMA.  A possible alternative could be to estimate the systemic velocity from the broad H$\alpha$ emission, although a rather long exposure time would be required in order to reach a suitably level of high signal to noise (c.f. Humphrey et al. 2008).

\section*{Acknowledgments}
AH thanks the Universidad Nacional Aut\'onomo de M\'exico and the Korea Astronomy and Space Science Institute for postdoctoral research fellowships.  LB was supported by the CONACyT grant J-50296.


\begin{thebibliography}{}

\bibitem[\protect\citeauthoryear{Anders 
\& Grevesse}{1989}]{1989GeCoA..53..197A} Anders E., Grevesse N., 1989, GeCoA, 53, 197 

\bibitem[\protect\citeauthoryear{Barkana \&
Loeb}{2003}]{2003Natur.421..341B} Barkana R., Loeb A., 2003, Natur, 421,
341

\bibitem[\protect\citeauthoryear{Barkana}{2004}]{2004MNRAS.347...59B}
Barkana R., 2004, MNRAS, 347, 59

\bibitem[\protect\citeauthoryear{Benson et al.}{2003}]{2003ApJ...599...38B}
Benson A.~J., Bower R.~G., Frenk C.~S., Lacey C.~G., Baugh C.~M., Cole S.,
2003, ApJ, 599, 38

\bibitem[\protect\citeauthoryear{Binette et
al.}{2000}]{2000A&A...356...23B} Binette L., Kurk J.~D.,
Villar-Mart{\'{\i}}n M., R{\"o}ttgering H.~J.~A., 2000, A\&A, 356, 23

\bibitem[\protect\citeauthoryear{Binette et
al.}{2006}]{2006A&A...459...31B} Binette L., Wilman R.~J.,
Villar-Mart{\'{\i}}n M., Fosbury R.~A.~E., Jarvis M.~J., R{\"o}ttgering
H.~J.~A., 2006, A\&A, 459, 31

\bibitem[\protect\citeauthoryear{Binette, Dopita, \&
Tuohy}{1985}]{1985ApJ...297..476B} Binette L., Dopita M.~A., Tuohy I.~R.,
1985, ApJ, 297, 476

\bibitem[\protect\citeauthoryear{Carilli et 
al.}{1997}]{1997ApJS..109....1C} Carilli C.~L., Roettgering H.~J.~A., van 
Ojik R., Miley G.~K., van Breugel W.~J.~M., 1997, ApJS, 109, 1 

\bibitem[\protect\citeauthoryear{Cimatti et 
al.}{1998}]{1998ApJ...499L..21C} Cimatti A., di Serego Alighieri S., Vernet 
J., Cohen M., Fosbury R.~A.~E., 1998, ApJ, 499, L21 

\bibitem[\protect\citeauthoryear{Conti, Leitherer, 
\& Vacca}{1996}]{1996ApJ...461L..87C} Conti P.~S., Leitherer C., Vacca W.~D., 1996, ApJ, 461, L87 

\bibitem[\protect\citeauthoryear{Cooke, Espey, \&
Carswell}{1997}]{1997MNRAS.284..552C} Cooke A.~J., Espey B., Carswell
R.~F., 1997, MNRAS, 284, 552

\bibitem[\protect\citeauthoryear{Crenshaw et 
al.}{1999}]{1999ApJ...516..750C} Crenshaw D.~M., Kraemer S.~B., Boggess A., 
Maran S.~P., Mushotzky R.~F., Wu C.-C., 1999, ApJ, 516, 750 

\bibitem[\protect\citeauthoryear{Daltabuit 
\& Cox}{1972}]{1972ApJ...173L..13D} Daltabuit E., Cox D., 1972, ApJ, 173, L13 

\bibitem[\protect\citeauthoryear{De Breuck et 
al.}{2003}]{2003A&A...401..911D} De Breuck C., et al., 2003, A\&A, 401, 911 

\bibitem[\protect\citeauthoryear{Dopita, Binette, 
\& Schwartz}{1982}]{1982ApJ...261..183D} Dopita M.~A., Binette L., Schwartz R.~D., 1982, ApJ, 261, 183 

\bibitem[\protect\citeauthoryear{Dopita 
\& Sutherland}{1995}]{1995ApJ...455..468D} Dopita M.~A., Sutherland R.~S., 1995, ApJ, 455, 468 

\bibitem[\protect\citeauthoryear{Dopita 
\& Sutherland}{1996}]{1996ApJS..102..161D} Dopita M.~A., Sutherland R.~S., 1996, ApJS, 102, 161 

\bibitem[\protect\citeauthoryear{Fardal, Giroux, \&
Shull}{1998}]{1998AJ....115.2206F} Fardal M.~A., Giroux M.~L., Shull J.~M.,
1998, AJ, 115, 2206

\bibitem[\protect\citeauthoryear{Ferruit et
al.}{1997}]{1997A&A...322...73F} Ferruit P., Binette L., Sutherland R.~S.,
Pecontal E., 1997, A\&A, 322, 73

\bibitem[\protect\citeauthoryear{Fosbury et
al.}{2003}]{2003ApJ...596..797F} Fosbury R.~A.~E., et al., 2003, ApJ, 596,
797

\bibitem[\protect\citeauthoryear{Goodrich, Cohen, 
\& Putney}{1995}]{1995PASP..107..179G} Goodrich R.~W., Cohen M.~H., Putney A., 1995, PASP, 107, 179 

\bibitem[\protect\citeauthoryear{Haiman, Spaans, \&
Quataert}{2000}]{2000ApJ...537L...5H} Haiman Z., Spaans M., Quataert E.,
2000, ApJ, 537, L5

\bibitem[\protect\citeauthoryear{Hamann 
\& Ferland}{1999}]{1999ARA&A..37..487H} Hamann F., Ferland G., 1999, ARA\&A, 37, 487 

\bibitem[\protect\citeauthoryear{Humphrey et 
al.}{2006}]{2006MNRAS.369.1103H} Humphrey A., Villar-Mart{\'{\i}}n M., 
Fosbury R., Vernet J., di Serego Alighieri S., 2006, MNRAS, 369, 1103 

\bibitem[\protect\citeauthoryear{Humphrey et
al.}{2007}]{2007MNRAS.375..705H} Humphrey A., Villar-Mart{\'{\i}}n M.,
Fosbury R., Binette L., Vernet J., De Breuck C., di Serego Alighieri S.,
2007a, MNRAS, 375, 705

\bibitem[\protect\citeauthoryear{Humphrey et 
al.}{2008}]{2008MNRAS.383...11H} Humphrey A., Villar-Mart{\'{\i}}n M., 
Vernet J., Fosbury R., di Serego Alighieri S., Binette L., 2008, MNRAS, 
383, 11 

\bibitem[\protect\citeauthoryear{Jarvis et al.}{2003}]{2003MNRAS.338..263J}
Jarvis M.~J., Wilman R.~J., R{\"o}ttgering H.~J.~A., Binette L., 2003,
MNRAS, 338, 263

\bibitem[\protect\citeauthoryear{Keenan et al.}{1985}]{1985MNRAS.214P..37K} 
Keenan F.~P., Johnson C.~T., Kingston A.~E., Dufton P.~L., 1985, MNRAS, 
214, 37P 

\bibitem[\protect\citeauthoryear{Keshet et al.}{2003}]{2003ApJ...585..128K}
Keshet U., Waxman E., Loeb A., Springel V., Hernquist L., 2003, ApJ, 585,
128

\bibitem[\protect\citeauthoryear{Korista, Ferland, 
\& Baldwin}{1997}]{1997ApJ...487..555K} Korista K., Ferland G., Baldwin J., 1997, ApJ, 487, 555 

\bibitem[\protect\citeauthoryear{Krause}{2002}]{2002A&A...386L...1K} Krause
M., 2002, A\&A, 386, L1

\bibitem[\protect\citeauthoryear{Kurk et al.}{2000}]{2000A&A...358L...1K}
Kurk J.~D., et al., 2000, A\&A, 358, L1

\bibitem[\protect\citeauthoryear{Larkin et al.}{2000}]{2000ApJ...533L..61L} 
Larkin J.~E., et al., 2000, ApJ, 533, L61 

\bibitem[\protect\citeauthoryear{LeFevre et
al.}{2003}]{2003SPIE.4841.1670L} LeFevre O., et al., 2003, SPIE, 4841, 1670

\bibitem[\protect\citeauthoryear{McCarthy et
al.}{1987}]{1987ApJ...319L..39M} McCarthy P.~J., Spinrad H., Djorgovski S.,
Strauss M.~A., van Breugel W., Liebert J., 1987, ApJ, 319, L39

\bibitem[\protect\citeauthoryear{McCarthy, Persson, 
\& West}{1992}]{1992ApJ...386...52M} McCarthy P.~J., Persson S.~E., West S.~C., 1992, ApJ, 386, 52 

\bibitem[\protect\citeauthoryear{Nesvadba et
al.}{2006}]{2006ApJ...650..693N} Nesvadba N.~P.~H., Lehnert M.~D.,
Eisenhauer F., Gilbert A., Tecza M., Abuter R., 2006, ApJ, 650, 693

\bibitem[\protect\citeauthoryear{Oke et al.}{1995}]{1995PASP..107..375O} 
Oke J.~B., et al., 1995, PASP, 107, 375 

\bibitem[\protect\citeauthoryear{Pentericci et 
al.}{1999}]{1999A&A...341..329P} Pentericci L., R{\"o}ttgering H.~J.~A., Miley G.~K., McCarthy P., Spinrad H., van Breugel W.~J.~M., Macchetto F., 1999, A\&A, 341, 329 

\bibitem[\protect\citeauthoryear{Pentericci et 
al.}{2001}]{2001ApJS..135...63P} Pentericci L., McCarthy P.~J., 
R{\"o}ttgering H.~J.~A., Miley G.~K., van Breugel W.~J.~M., Fosbury R., 
2001, ApJS, 135, 63 

\bibitem[\protect\citeauthoryear{Pettini et 
al.}{2000}]{2000ApJ...528...96P} Pettini M., Steidel C.~C., Adelberger 
K.~L., Dickinson M., Giavalisco M., 2000, ApJ, 528, 96 

\bibitem[\protect\citeauthoryear{Rocca-Volmerange et
al.}{2004}]{2004A&A...415..931R} Rocca-Volmerange B., Le Borgne D., De
Breuck C., Fioc M., Moy E., 2004, A\&A, 415, 931

\bibitem[\protect\citeauthoryear{Rottgering et
al.}{1995}]{1995MNRAS.277..389R} R\"ottgering H.~J.~A., Hunstead R.~W., Miley
G.~K., van Ojik R., Wieringa M.~H., 1995, MNRAS, 277, 389

\bibitem[\protect\citeauthoryear{R\"ottgering et
al.}{1997}]{1997A&A...326..505R} R\"ottgering H.~J.~A., van Ojik R., Miley
G.~K., Chambers K.~C., van Breugel W.~J.~M., de Koff S., 1997, A\&A, 326,
505

\bibitem[\protect\citeauthoryear{S{\'a}nchez}{2006}]{2006AN....327..850S} S{\'a}nchez S.~F., 2006, AN, 327, 850 

\bibitem[\protect\citeauthoryear{Schaye et al.}{2003}]{2003ApJ...596..768S} 
Schaye J., Aguirre A., Kim T.-S., Theuns T., Rauch M., Sargent W.~L.~W., 
2003, ApJ, 596, 768 

\bibitem[\protect\citeauthoryear{Seymour et
al.}{2007}]{2007ApJS..171..353S} Seymour N., et al., 2007, ApJS, 171, 353

\bibitem[\protect\citeauthoryear{Spitzer}{1978}{}]
Spitzer L., 1978, Physical Processes in the Interstellar Medium. John Wiley \& Sons, New York.  

\bibitem[\protect\citeauthoryear{van Ojik et
al.}{1997}]{1997A&A...317..358V} van Ojik R., R\"ottgering H.~J.~A., Miley
G.~K., Hunstead R.~W., 1997, A\&A, 317, 358

\bibitem[\protect\citeauthoryear{Vernet et 
al.}{2001}]{2001A&A...366....7V} Vernet J., Fosbury R.~A.~E., Villar-Mart{\'{\i}}n M., Cohen M.~H., Cimatti A., di Serego Alighieri S., Goodrich R.~W., 2001, A\&A, 366, 7 

\bibitem[\protect\citeauthoryear{Villar-Mart{\'{\i}}n et
al.}{1999}]{1999A&A...351...47V} Villar-Mart{\'{\i}}n M., Fosbury R.~A.~E.,
Binette L., Tadhunter C.~N., Rocca-Volmerange B., 1999, A\&A, 351, 47

\bibitem[\protect\citeauthoryear{Villar-Mart{\'{\i}}n et
al.}{2003}]{2003MNRAS.346..273V} Villar-Mart{\'{\i}}n M., Vernet J., di
Serego Alighieri S., Fosbury R., Humphrey A., Pentericci L., 2003, MNRAS,
346, 273

\bibitem[\protect\citeauthoryear{Villar-Mart{\'{\i}}n et
al.}{2007}]{2007MNRAS.378..416V} Villar-Mart{\'{\i}}n M., S{\'a}nchez
S.~F., Humphrey A., Dijkstra M., di Serego Alighieri S., De Breuck C.,
Gonz{\'a}lez Delgado R., 2007, MNRAS, 378, 416

\bibitem[\protect\citeauthoryear{Weidinger, M{\o}ller, \&
Fynbo}{2004}]{2004Natur.430..999W} Weidinger M., M{\o}ller P., Fynbo
J.~P.~U., 2004, Natur, 430, 999

\bibitem[\protect\citeauthoryear{Wilman et al.}{2005}]{2005Natur.436..227W}
Wilman R.~J., Gerssen J., Bower R.~G., Morris S.~L., Bacon R., de Zeeuw
P.~T., Davies R.~L., 2005, Natur, 436, 227

\bibitem[\protect\citeauthoryear{Wilman et al.}{2004}]{2004MNRAS.351.1109W}
Wilman R.~J., Jarvis M.~J., R{\"o}ttgering H.~J.~A., Binette L., 2004,
MNRAS, 351, 1109

\bibitem[\protect\citeauthoryear{Zirm et al.}{2005}]{2005ApJ...630...68Z} 
Zirm A.~W., et al., 2005, ApJ, 630, 68 

\end{thebibliography}
\end{document}